\documentclass[iop,numberedappendix,appendixfloats]{emulateapj}

\usepackage[parfill]{parskip}    
\usepackage{graphicx}
\usepackage{amssymb}
\usepackage{epstopdf}
\usepackage{amsmath} 
\usepackage{amsthm} 
\usepackage{amssymb}	
\usepackage{graphicx} 
\usepackage{bm}
\usepackage{enumerate}
\def\apj{ApJ}

\def\sextractor{{\sc SExtractor }}

\def\eg{{\it e.g.\ }}

\newcommand{\bs}{\boldsymbol}

\def\sun{\odot}

\def\mz{z_{850}} 
\def\mmax{m_{\rm max}}

\def\numobs{N^{\rm obs}}

\def\data{\bs{d}}
\def\allDat{\bs{D}}

\def\pos{\bs{x}}

\def\epss{\epsilon_{\rm s}}
\def\epsh{\epsilon_{\rm h}}

\def\epss{\epsilon_{\rm s}}

\def\rpower{\gamma_{\rm p}}

\def\Numsat{N_{\rm{s} }}

\def\z8{$z_{850}$}
\def\z{z$_{850}$}

\def\epsh{\epsilon_{\rm h}}

\def\hdat{\bs{h}}
\def\numhosts{N_{\rm h}}

\def\Rhost{R_{\rm h}}

\def\prob{{\rm Pr}}

\def\be{\begin{equation}}
\def\ee{\end{equation}}

\def\dm{\Delta m}

\def\numobjs{N}
\def\nobjs{N}

\def\modpars{\bs{\theta}}

\def\nobs{N^{\rm obs}}
\def\numsat{N_{\rm s}}

\slugcomment{Accepted for publication in \apj}

\shortauthors{Nierenberg et al.}
\shorttitle{Satellite Spatial Distribution}

\begin{document}

\title{Luminous satellites of early-type galaxies I: Spatial distribution}

\author{A.M. Nierenberg\altaffilmark{1}$^{*}$, M.W. Auger\altaffilmark{1}, T. Treu\altaffilmark{1}, P.J. Marshall\altaffilmark{2,3}, C.D. Fassnacht\altaffilmark{4}}
\altaffiltext{1}{Department of Physics, University of California, Santa Barbara, CA 93106, USA}
\altaffiltext{2}{KIPAC, Stanford University, Stanford, CA 94306, USA}
\altaffiltext{3}{Department of Physics, University of Oxford, Keble Road, Oxford, OX1 3RH, UK}
\altaffiltext{4}{Department of Physics, University of California, Davis, CA 95616, USA}
\altaffiltext{*}{{\tt amn01@physics.ucsb.edu}}

\begin{abstract}
We study the spatial distribution of faint satellites of intermediate redshift ($0.1<z<0.8$),
early-type galaxies, selected
from the GOODS fields. We combine high resolution HST images and
state-of-the-art host subtraction techniques to detect satellites of
unprecedented faintness and proximity to intermediate redshift host
galaxies (up to 5.5 magnitudes fainter and as close as
0\farcs5/2.5~kpc to the host centers). 
We model the spatial distribution of objects near the hosts as a combination of an isotropic, homogeneous background/foreground population and a satellite population with a power law radial profile and an elliptical angular distribution. We detect a significant population of
satellites ($\Numsat=1.7^{+0.9}_{-0.8}$) that is comparable to the number of Milky Way satellites with similar host-satellite contrast. 
The average projected radial profile 
of the satellite distribution is isothermal ($\rpower=-1.0^{+0.3}_{-0.4}$), which is consistent with the observed central mass density profile of massive early-type galaxies. Furthermore, the satellite distribution is highly anisotropic (isotropy is ruled out at a $>$99.99\% confidence level).  Defining $\phi$ to be the offset between the major axis of the
satellite spatial distribution and the major axis of the host light profile, we 
find a maximum posterior probability of  
$\phi = 0$ and $|\phi|$ less than $42^{o}$ at the
68\% confidence level. The alignment of
the satellite distribution with the light of the host is consistent
with simulations, assuming that light traces mass for the host galaxy
as observed for lens galaxies. The anisotropy of the
satellite population enhances its ability to produce the flux ratio
anomalies observed in gravitationally lensed quasars.
\end{abstract}

 ---------------------------------------------------------------------------
                      
\section{Introduction}
\label{sec:intro}

Cold Dark Matter (CDM) simulations have been successful at predicting
the distribution of matter on super-galaxy scales (greater than a few
Mpc), but less successful on smaller scales, predicting orders of
magnitude more satellite galaxies than are observed in the local group
\citep{Strigari++07}.  The solution to the so-called missing satellite
problem will have important implications for dark matter cosmology and
star formation in low mass halos \citep{Klypin++1999,Moore++1999}.
The discrepancy between observed and predicted satellite numbers could
be due to incorrect assumptions about the nature of dark matter in the
cosmological simulations.  For instance, it has been proposed that
small scale structure does not form because dark matter is not
actually ``cold'' but instead has appreciable velocity
\citep[e.g.][]{Colin++00}, or that the power spectrum of density fluctuations used in simulations is incorrect \citep[e.g.][]{Kam++00,Zen++03}. Alternatively,
it is possible that the simulations are correct but that very low
mass satellites are too faint to be detected. The luminosity of
satellites is difficult to predict theoretically as high resolution
CDM simulations do not include baryonic interactions and therefore
neglect how processes such as UV reionization, baryonic cooling,
supernovae, and gas accretion affect star formation in dark matter halos
\citep{Thoul++1996,Gnedin++00,Kaufmann++08}.  Hydrodynamical
simulations which include gas and radiative transfer have been able to
reproduce some aspects of the observed luminosity and metallicity
properties of galaxies \citep[see ][and references
therein]{Springel++10}.  However, these simulations must be improved
in order to attain the complete and self-consistent understanding of
star formation which is needed to predict the luminosity function of
satellite galaxies.

While the luminosity and mass functions of satellites can be used to
constrain cosmology and star formation, their spatial distribution can
be used as a tracer of the total mass distribution of galaxy halos. Numerical simulations \citep{Krav++10} predict that satellites should
be distributed as the mass profile of the host galaxy which,
neglecting baryons, is expected to be approximately a
\citet[][hereafter NFW]{NFW++1996} profile. However, the inner region of the host halo profile
is likely to be steeper in the presence of baryons \citep[e.g.,][]{Blu++86,Gne++04}. From gravitational lensing techniques, the total central mass density distributions of massive early-type galaxies have been shown to follow approximately isothermal profiles \citep[i.e., $\rho(r)\propto r^{-2}$;][]{Cohn++01,Treu++04, Koopmans++06,Koopmans++09, Aug++10}. The two point correlation function of luminous red galaxies (limiting
magnitude $\sim L^*$) in the Sloan Digital Sky Survey (SDSS) has been
shown to be consistent with an isothermal satellite
distribution \citep[][hereafter W10]{Watson++10}, although
\citet{Chen++08} (hereafter C08) studied fainter satellites with a
limiting magnitude of about 0.1$L^{*}$ and found that these satellites
followed a distribution closer to NFW. This apparent discrepancy in the form of the radial profile of satellites is likely due to the differences in how line-of-sight interlopers are treated in the two studies and the different radial ranges that were probed. The angular profile of satellites is also expected to follow the host mass distribution, and simulations predict that satellites will have an anisotropic spatial distribution, appearing preferentially along the
major axis of the host mass profile \citep[e.g.][]{Zentner++05,
Zentner++06, Knebe++04,Aubert++04,Faltenbacher++07,Faltenbacher++08}. This anisotropy has been observed in the local group ~\citep[e.g.][]{Metz++09} and in SDSS galaxy distributions \citep{Brainerd++05,Yang++06,Wang++08,Agustsson++10}.

Furthermore, the abundance of bright satellites (typically brighter
than about 0.1 $L^*$) near hosts, also known as close pairs, has been
used to constrain the rate of major mergers given assumptions about
merger timescales and stellar mass to virial mass ratios
\citep[e.g.][]{Bell++06, Patton++08, Bundy++09, Robaina++10,
LeFevre++00}. Mergers are believed to play a key role in galaxy growth.
For instance, simulations by \citet{Hopkins++10} indicate
that about 30 percent of the mass in galaxy bulges comes from minor
mergers (mergers where the mass ratio between merging objects is less
than 1/3). Furthermore, observations and simulations have shown that minor mergers may play a dominant role in determining the recent (z$<1$) star formation activity in early-type galaxies \citep{Kaviraj++09,Kaviraj++11}.
One would therefore like to study the merger rates at
lower virial and stellar mass scales \citep[see also][]{Bundy++07,
Naab++09, Fakhouri++10}, although this requires finding and studying
faint satellites at cosmological distances.

\citet[][hereafter J10]{Jackson++10} have used high resolution Hubble
Space Telescope (HST) imaging from the COSMOS survey \citep{Scov++07}
in order to search for satellites around approximately 11,000
massive galaxies. They searched for companion
objects within 2\arcsec~ from their hosts, although the closest objects remained
undetected as can be seen, for example, in the third panel of Figure 3
of their paper.  They compared their findings to the number of
satellites that have been found around 22 CLASS lens systems and found that
the lens systems were more likely to have a companion object than a
typical COSMOS elliptical galaxy.

The efforts of \citet{Jackson++10} highlight the difficulties
 of studying satellites at higher
redshifts, where decreasing physical resolution reduces the number
of faint companions that can be observed near bright hosts.
However, one successful method of studying the mass and positional properties
of high redshift satellites has been observations and simulations of
gravitationally lensed active galactic nuclei (AGN)
\citep[e.g.][]{Dalal++02,McK++05, Vegetti++09, Xu++09}. This method
shows great promise with regards to the missing satellite problem
because the brightness of the lensed images is very sensitive to local
perturbations of the deflector potential, such as those expected from
substructure. Thus, mass models which reproduce lensed image magnitudes
can be used to infer the presence of satellites regardless of whether
or not the satellites are luminous. However, there are two main
difficulties with using lensed quasars to study satellites. The first
is the small number of suitable strong lens systems currently known
and the complexity of their selection function \citep[e.g.,][and
references therein]{Treu++10}. This limitation will become less
important with new surveys such as the Large Synoptic Survey Telescope
(LSST) which will vastly increase the number of detected quasar lenses
\citep{Oguri++10}. A second and more significant difficulty with using
lensing to study satellite properties is that bright lensed images
make it difficult or impossible to measure the luminosity of
perturbing satellites \citep[e.g.][]{McK++05}, although this will
be mitigated by future telescopes with higher resolution than what is currently
available. In addition to AGN, recent work has shown that dark satellites can be detected by reconstructing the lensing potential
using extended background sources
\citep[e.g.,][]{Koo05,Vegetti++09,Vegetti++10}, further enlarging the
sample of gravitational lenses that can be used to detect
substructure.  Furthermore, substructure might be detectable
using detailed positional and time-delay
measurements of lensed images rather than just magnification
\citep{MacLeod++09} thus further addressing the first
limitation.

A powerful approach to understanding star formation efficiency in low
mass satellite galaxies is the combination of lensing studies to
constrain the mass function of satellites, and imaging studies to
constrain the luminosity function \citep[][and references
therein]{Treu++10, Krav++10}. With this goal in mind, we have started a new program to characterize
the visible properties of faint satellites of massive galaxies. In
this first paper, we focus on the spatial distribution of faint
satellites of early-type galaxies at intermediate redshifts,
$0.1<z<0.8$, selected from the GOODS fields \citep{Gia++04}.  We
concentrate on this population of hosts because early-type galaxies
dominate the sample of strong lensing galaxies \citep{Aug++09} and are
therefore the proper host sample for comparison to the satellite mass
function results from lensing studies.  An additional benefit to
studying early-type galaxies is that they have relatively smooth surface
brightness profiles which are ideal when searching for nearby compact
and faint companions.

For this initial analysis we only require one photometric band. We
use $\mz$ because it is the reddest of the GOODS bands and therefore
it is the most faithful tracer of stellar mass. In forthcoming papers
of this series we will use multiple band photometry to improve
line-of-sight interloper removal and constrain the luminosity
and stellar mass functions of the satellites (Nierenberg et al. 2011a,
in preparation; hereafter paper II), and we will combine our results
with the total mass function obtained from strong
lensing studies to constrain the total-to-stellar mass ratio of the
satellites (Nierenberg et al. 2011b, in preparation; hereafter paper
III).

This paper is organized as follows: In \S \ref{sec:sample} we discuss
the properties of our host galaxy sample. In \S \ref{sec:subtraction}
we summarize our image analysis, including our elliptical B-spline
host galaxy subtraction and faint object detection
and photometry methodologies. In \S \ref{sec:firstlook} we take a first
look at the satellite distribution by means of a binned analysis,
which is useful for visualizing the main trends and identifying the
strength of the signal. In \S \ref{sec:model} we describe our model
for the combined satellite plus background object spatial distribution and
the parameters that we aim to constrain (average number of satellites,
power law slope, etc.) along with a number of nuisance parameters
(density of the background population, slope of the background number
counts, etc.). In \S~\ref{sec:results} we present the results obtained
by comparing our model to the data. In \S~\ref{sec:discussion} we
discuss our results and compare them to previous satellite studies.
In \S~\ref{sec:summary} we provide a concise summary. The appendix
contains more detailed explanations of many of the methods we used in
this paper.

Throughout this paper, we assume a flat $\Lambda$CDM cosmology with
$h=0.7$ and $\Omega_{\rm m}=0.3$.  All magnitudes are given in the AB
system \citep{Oke++1974} unless otherwise stated.

                      
\section{Host Galaxy Sample}
\label{sec:sample}
We select a population of early-type (E and S0) host galaxies from the catalog of \citet{Bun++05_cat}, which contains
spectroscopic redshifts,stellar mass estimates, and morphological classifications for 
47\% of its objects \citep[see][]{Treu++05}. The objects in this catalog were originally
selected from Hubble Space Telescope photometric catalogs \footnote{The GOODS catalogs are 
available at \texttt{http://archive.stsci.edu/prepds/goods}} made by the 
GOODS team using the \sextractor software \citep{Bertin++1996}. In the GOODS-South field, we use COMBO-17 photometric redshifts \citep{Wolf++04} for our hosts
where spectroscopic redshifts are not
available.  In the GOODS-North field only a handful of objects have morphological classifications and 
stellar mass estimates but no spectroscopic redshifts. We excluded these from
our analysis.

We limit the host redshifts to $z<0.8$ to guarantee that we can detect
satellites with host-satellite luminosity contrasts fainter than the equivalent
contrast between the Small Magellanic Cloud and the Milky Way at all redshifts, thus
increasing the likelihood that we observe approximately one satellite per
host. We also exclude $z<0.1$ galaxies, which are few (owing to the
small volume) and too extended in angular size to analyze in the same
way as the more distant sample.  Finally we exclude two hosts 
which appear to be undergoing major mergers as these have 
physical environments which are significantly distinct from the majority of our sample.
To ensure that we study the same population of satellites for all
hosts, despite their broad distribution in redshift, we only consider
satellites brighter than a fixed fraction ($\dm$) of the host galaxy
luminosity (i.e. a fixed difference in magnitude). 
We also require that all objects be brighter than the detection
threshold (\z $=\mmax=26.5$, as described in
Section~\ref{subsec:objectDetection}), and therefore we cut the parent
sample to a maximum value of \z, depending on the choice of $\dm$, such
that $m_{\rm host} + \dm < \mmax$. 

\begin{figure}[h!]
\centering
\includegraphics[scale=0.4]{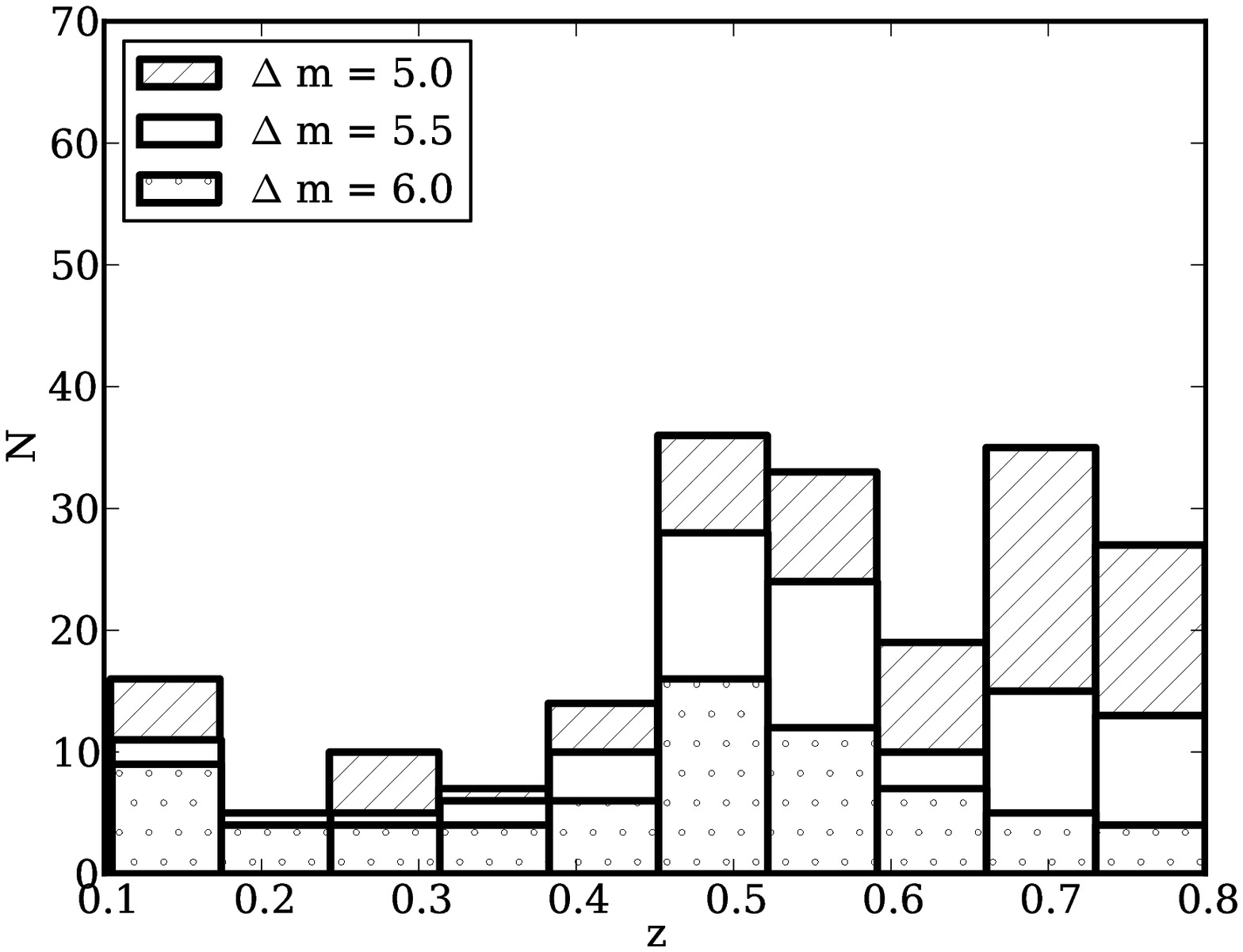} 
\includegraphics[scale=0.4]{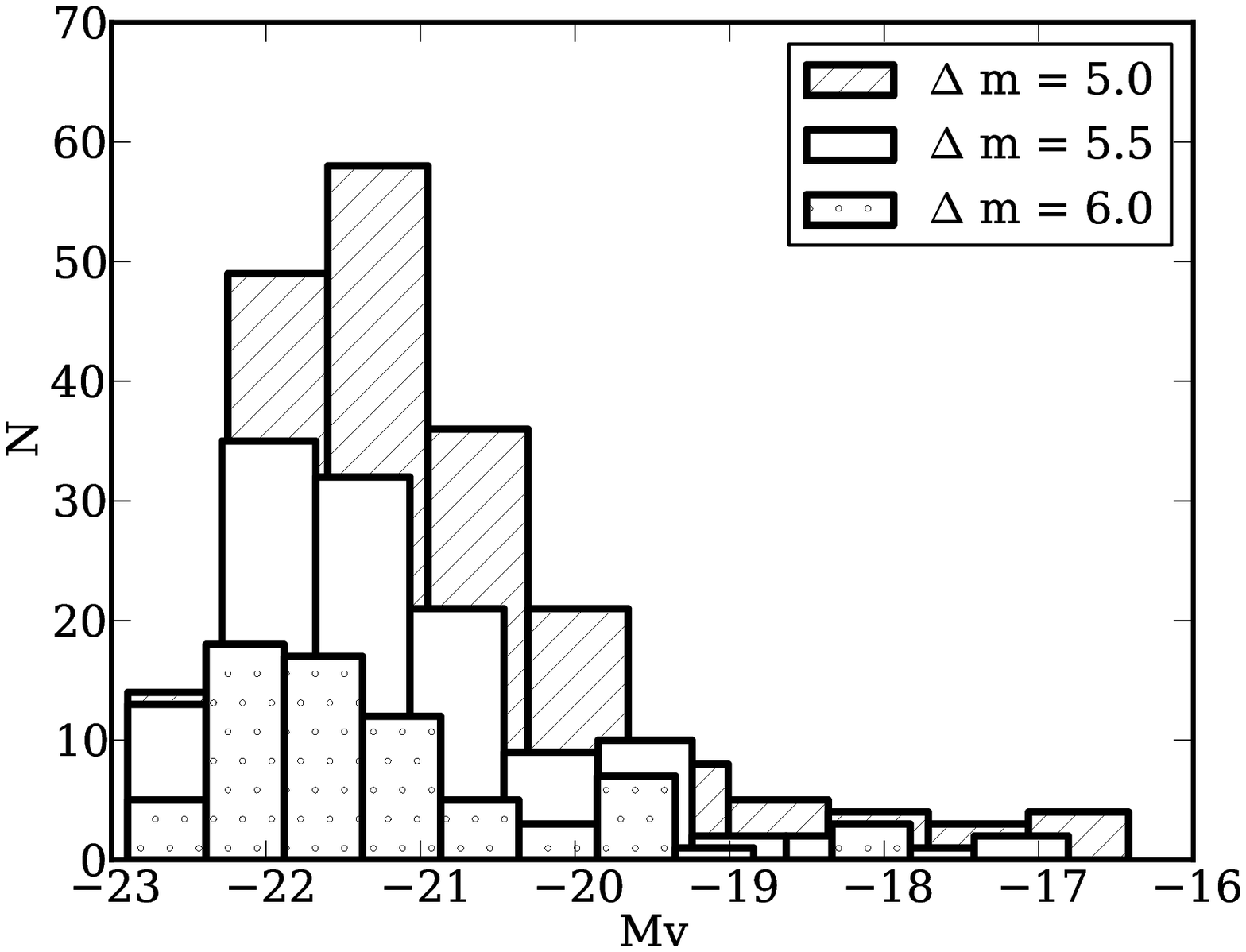} 
\includegraphics[scale=0.4]{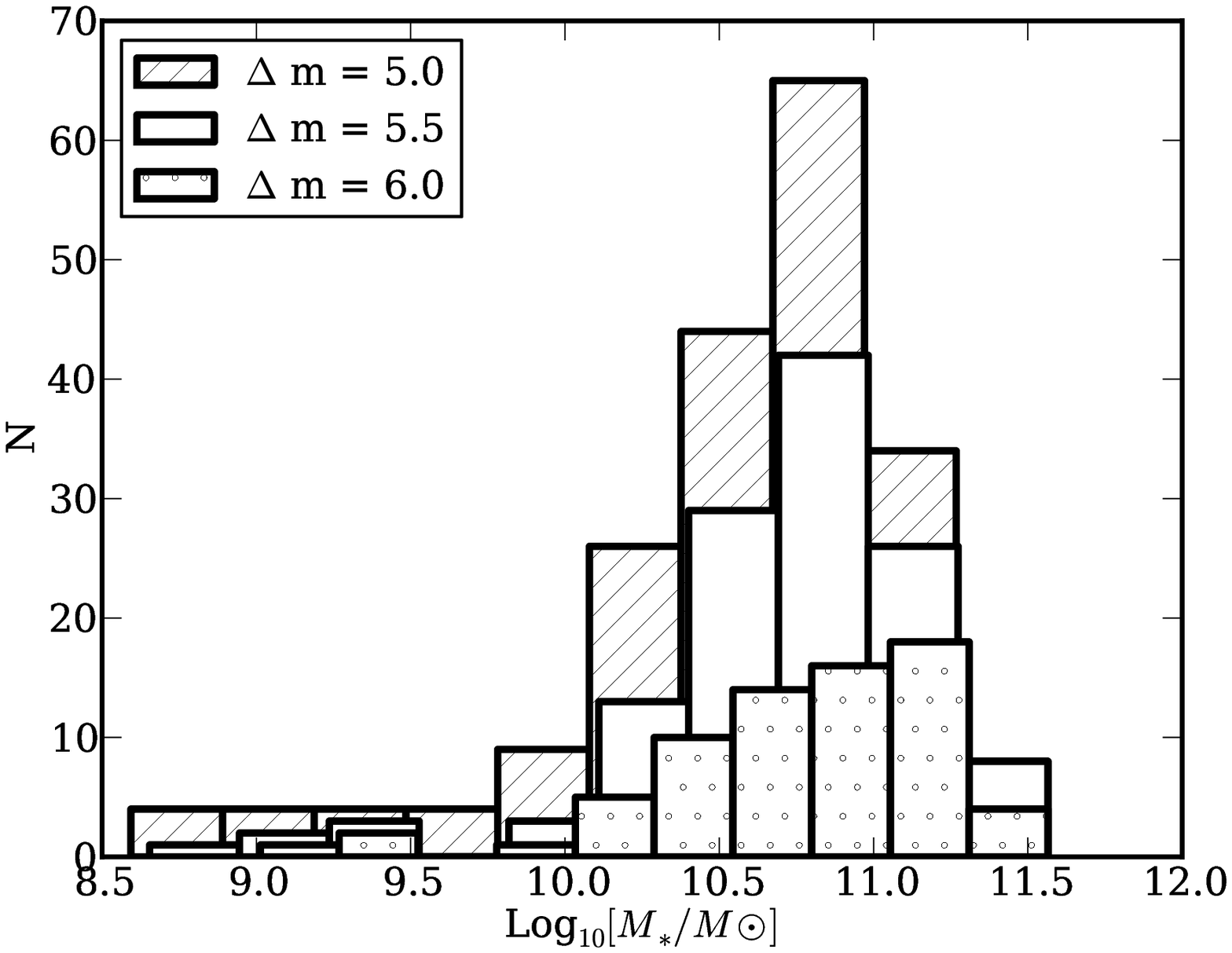} 
\caption{From top to bottom: The distribution of host redshifts, absolute V band magnitudes, and stellar masses in three
host completeness ranges ($\dm < \mmax - m_{\rm host}$) taken from \citet{Bun++05_cat}.}
\label{fig:host_redshifts}
\end{figure}

As we increase the size of the magnitude range we study, the number of
hosts that are complete within that magnitude range drops; there are
202, 127 and 71 hosts complete to $\dm =$ 6.0, 5.5, and 5.0
respectively in the final host sample. At the same time, the number of satellites per host is
expected to increase with $\dm$. The optimal choice of $\dm$ needs to
strike a balance between the two effects. As we will show, for the present GOODS dataset we find that $\dm$=5.5 maximizes
the signal to noise ratio of the detection and therefore we will adopt
this choice as our default. To illustrate the robustness of our
results to small changes in $\dm$, we will also describe our findings
for $\dm$ = 5 and 6.  The distributions of host redshifts, absolute
magnitudes, and stellar masses for the three choices of $\dm$ are shown
in Figure~\ref{fig:host_redshifts}. As $\dm$ becomes larger, the hosts
that satisfy the completeness requirements tend to become brighter and
shift to lower redshift.  As with any flux limited sample, the lower
redshift objects of the sample will be dominated by the more abundant,
intrinsically fainter galaxies, while the higher redshift objects will
be dominated by intrinsically brighter objects. We defer an analysis
of evolutionary trends to papers II and III where we will also study
the luminosity and stellar mass properties of the satellites. Results
in this paper are an average of the properties of the satellite
population in the $0.1<z<0.8$ redshift range.

We use the second order moment of the host intensity distribution, measured along the major axis by \sextractor\footnote{\sextractor's AWIN\_IMAGE} as a scale factor which we represent by $\Rhost$.
$\Rhost$ compensates not only for
varying angular size with redshift, but is also intrinsically related
to host masses via the size-mass relation
\citep[e.g.][]{Tru++06,Williams++10} and thus adjusts for variations
in host masses at a given redshift. $\Rhost$ varies across the host sample
in physical size (0.8-5.5 kpc) with host mass variations and in
angular size (between 0\farcs2-1\farcs1).  The median (and modal) values of $\Rhost$ are 0\farcs5 and 3 kpc. We use
these as fiducial values when converting our results to angular and
physical scales.

\section{Detection and photometry of close neighbors}
\label{sec:subtraction}

Companions of high redshift galaxies are difficult to study because
they are intrinsically faint and often obscured by the host galaxy
light.  In this section we describe our method of modeling and
subtracting the host galaxy light profile to allow us to identify and
perform accurate photometry on nearby objects.  Note that while we
model host light in small regions around each host, we do not limit
our analysis to objects within this modeled region. We use the GOODS
catalogs for photometry and astrometry for all objects outside of the
modeled cutout region. 

\subsection{Host Galaxy B-Spline Models}
Producing a good model for the light profile of the host galaxies is critical
for studying the population of faint objects near the host. We use a 
multi-step process to create an accurate model of the host light profile
which does not include the light from nearby objects.

We model the host light profile in a small cutout centered on the host
galaxy.  We choose this cutout to be 20 $\times$ 20 $\Rhost$ in order to ensure that all
significant levels of host light are removed. In the first step of the modeling process, we use \sextractor to
identify all objects in the cutout region. The \sextractor parameters
at this stage are chosen to err on the side of identifying noise peaks as objects in
order to ensure that all real objects being obscured by the host light are
recognized; see Appendix~\ref{app:subtraction} for a detailed
description of the \sextractor parameters and how they were
chosen. \sextractor outputs a segmentation map which we use to mask
out all identified objects other than the host galaxy.  \sextractor
also returns measurements of the axis ratio and position angle of the
host galaxy which we use in addition to $\Rhost$ to define an
elliptical coordinate system for the cutout.
 
We then model the host light in the masked image using empirical polar
B-spline models.  We choose B-spline models because they quickly fit
the observed light distribution with a smooth model that is
independent of PSF effects and is more flexible than Sersic models,
for example. Our method is similar to that described by
\citet{Bol++05,Bol++06}. Each iteration of the B-spline code fits a
model to the masked data, subtracts the model, and then identifies and
masks new residual structures. This process is repeated three times.
The final B-spline model is then subtracted from the data to produce a
residual image, which is used to perform object detection and
photometry for field galaxies near the host.  Further details of the
modeling and masking procedure are provided in Appendix
\ref{app:subtraction}. Examples of our host subtraction procedure are
shown in Figure~\ref{fig:modExamples}.

\subsection{Object Detection and Photometry}
\label{subsec:objectDetection}

We use \sextractor to detect and measure the properties of objects
within the host-subtracted cutouts. To ensure completeness, we limit
the sample to objects with MAG\_AUTO \z$<26.5$ magnitudes, where the
GOODS images are virtually 100\% complete, even close to the host
galaxy as we will show below. Furthermore, we only study objects
fainter than \z$>21.0$ magnitudes. This is because our analysis relies
on an accurate characterization of the background (see Section
\ref{sec:backgroundmodel}) and very bright objects appear in low
numbers with large fluctuations which can bias the number counts slope
near a particular host. 

Because we use the GOODS catalog data outside of our cutouts, it is
imperative that our detections and photometry after host light
subtraction are as close as possible to GOODS detections and
photometry.  We confirmed this by running \sextractor with our
parameters on a large un-modeled section of the GOODS field. We
recovered virtually the same number of objects and our photometry was
consistent with GOODS photometry.  This comparison is described in
further detail in Appendix ~\ref{app:SE}.

Due to the relative brightness of hosts compared to satellites, there
is a minimum radius at which we will be able to accurately identify
faint objects, regardless of how careful we are during the host
modeling process.  

By simulating point sources at and fainter
than our chosen limiting magnitude (26.5) at varying distances from a
representative subset of hosts, we find the minimum radius for
completeness to be 1.5 $\Rhost$ ($\sim$ 0\farcs9) for the vast majority of
cases. The few cases for which highly flattened hosts have significant
residuals extending to a larger distance are identified manually; for
those systems, appropriately larger inner regions are excluded from
our analysis (see Appendix \ref{app:subtraction}). 

Note that in the redshift range we study, even the most intrinsically
faint sources will have effective radii typically less than 0\farcs1 \citep[see, e.g.,][]{deRijcke++09} 
which is smaller than the FWHM of the GOODS PSF. Thus these sources
are effectively point sources. We test our sensitivity to intrinsically faint sources 
by simulating faint exponential disks near our hosts
at varying redshifts, with effective 
radii estimated from the relations given by \citet{deRijcke++09}. 
In the innermost region, between 1.5 and 3.5 $\Rhost$,  
we failed to detect approximately 20\% \footnote{The true incompleteness
estimate requires knowledge of the object luminosity function. We make a rough estimate by averaging the 
results for objects at redshifts of 0.1, 0.4 and 0.8, where the redshift 0.1 objects
are the most intrinsically faint and most difficult to detect.} 
of 26.5 magnitude objects. 
Outside of this region we recovered approximately 100\% of the simulated objects. 
For brighter objects with apparent magnitudes
of 24.5 our recovery was approximately 100\% complete even in this inner region.  
In our final analysis we study the satellite population between $1.5$ and 45 $\Rhost$. 
Taking this into account, we can examine the worst case scenario in which all satellites 
are exponential disks with apparent magnitudes of 26.5. 
Assuming that satellites are distributed radially in projection as 
$P_{sat}(r) \propto r^{-1}$, 
we would only underestimate the final satellite number by 1-2 \% 
because even in this extreme case, the incompleteness is confined to a small region. 
Thus we expect that the loss of completeness in the innermost region for the  
faintest sources will have a negligible effect on the analysis of the total sample.

\begin{figure*}
\centering
\includegraphics[scale= 0.6]{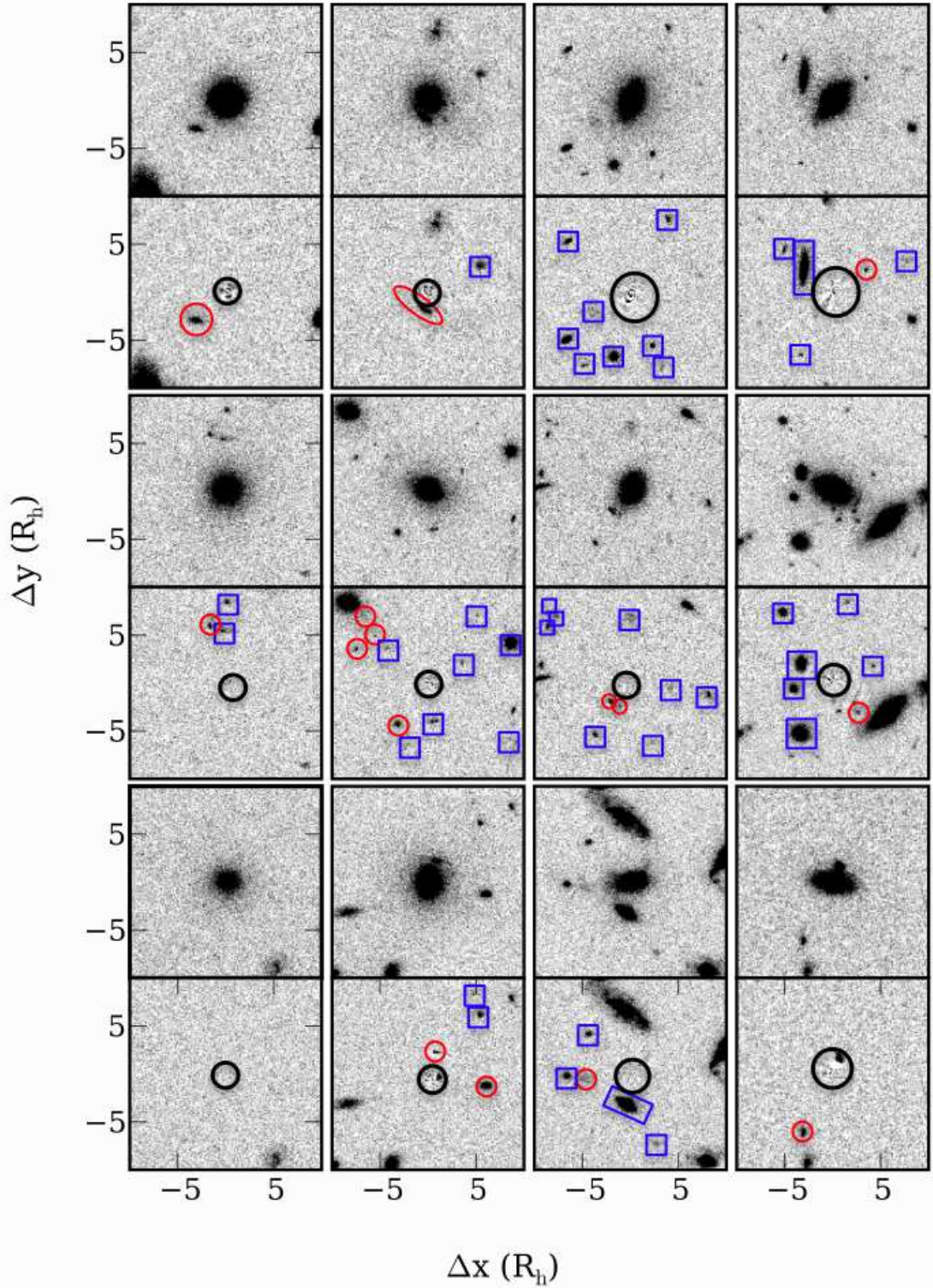} 
\caption{Demonstration of galaxy modeling for a range of
host ellipticities and sizes. Original images are shown with residuals
after host model subtraction immediately below.  Host galaxy
physical size and ellipticity decrease downwards and to the left. All
images are 20 $\Rhost$ on a side. New object detections, made possible
by the host light subtraction, are circled in red. Objects that were
detected in the GOODS catalogs and that have complete photometry
within the cutout (i.e. did not raise a \sextractor flag greater than
2) are identified by blue squares. Some objects visible to the eye
were omitted from our final catalog because they were fainter than our
detection limit (for instance, the object just outside of the excluded region in the top right galaxy). 
The central region of the host, excluded in our
analysis, is identified by a black circle.}
\label{fig:modExamples}
\end{figure*}

To compile a single catalog and avoid duplications, we compare the
position of each object identified in the residual images to objects
already in the GOODS catalogs. If the object is already in the
catalogs (position within 0\farcs3), we replace the GOODS photometry
and astrometry with our own measurements, which do not suffer from
being contaminated by host-galaxy light (see Figure
\ref{fig:mevgoodsmagauto}). The mean distance between `matched' objects
was $4 \pm 3$ pixels, this corresponds to 0\farcs12 which is the FWHM of the
GOODS PSF. For objects that are not matched to
objects already in the GOODS catalog, we add a new entry. Finally, for
all objects outside of the host-subtracted cutouts, we use the
measurements from the GOODS catalog and we do not attempt to detect
new objects. In Figure \ref{fig:modExamples} we show examples of the modeling
process for a variety of host ellipticities and physical sizes, along
with newly detected objects in the host-subtracted images.

\subsection{Objects Detected in Cutout Regions}
\label{subsec:newObjects}
\begin{figure}[h!]
\centering
\includegraphics[scale=0.45]{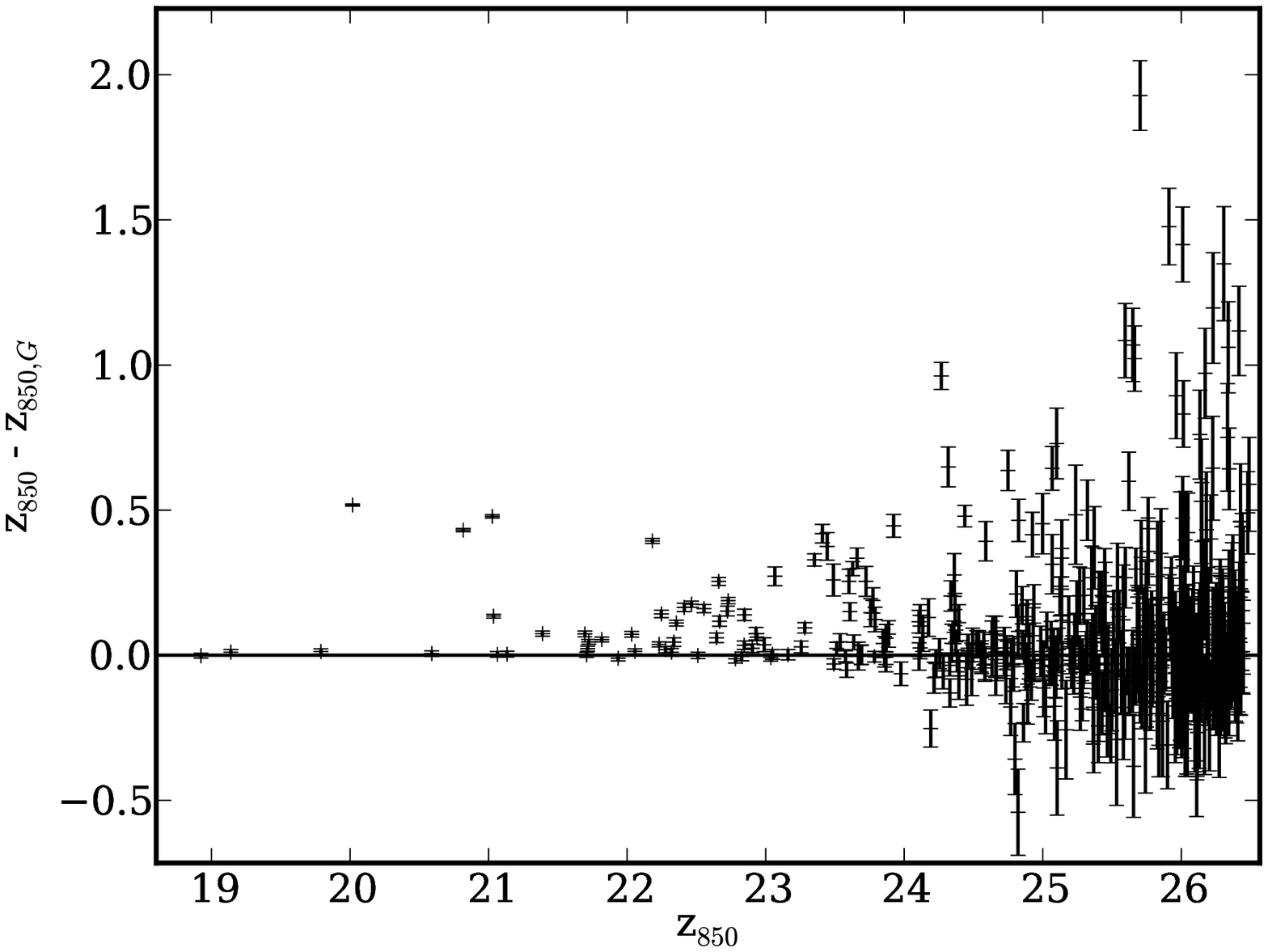} 
\caption{Comparison of photometry before (z$_{850,G}$) and after ($\mz$) host subtraction for objects in cutout regions
which had already been identified in the GOODS catalogs prior to host subtraction.}
\label{fig:mevgoodsmagauto}
\end{figure}

The host light subtraction has two important effects on our measurement
of objects near the host galaxy. The first is that it removes host light
contamination and allows for accurate photometry of nearby objects, 
as shown in Figure~\ref{fig:mevgoodsmagauto}. This figure compares
GOODS photometry to host-subtracted photometry
for objects that had already been identified in the GOODS catalogs.
As expected, we measure object
magnitudes to be slightly fainter than the GOODS photometry after host
subtraction, with a mean difference of 0.12 $\pm$ 0.02 magnitudes for objects
within 8 $\Rhost$ ($\sim$ 4\arcsec) of the host galaxy. Our photometry is identical 
to GOODS photometry without host subtraction
(see Appendix \ref{app:SE}), and we confirmed that the host light 
removal was accurate by simulating faint point sources near the hosts and ensuring 
that they were recovered with accurate photometry. Thus
the difference in magnitude after host subtraction shown in Figure~\ref{fig:mevgoodsmagauto}  
is entirely due to the removal of the host light contamination which has a significant impact on photometry
performed near bright objects.

The second important effect of host light subtraction is that it 
allows us to detect new objects. If the newly detected objects are real 
rather than artifacts of the host light subtraction,
 then we expect that object properties such as
brightness and elongation will be similar to the properties of objects
that had already been detected in the GOODS fields.  
The effect of weak lensing on magnification and shear is negligible for the small number of 
background sources close to the host Einstein radii in projection (the affected region is
typically an annulus of $\sim$ 1\farcs0 $\pm$ 0\farcs2 for 
massive ellipticals). One way to see this is using the fact that the effect of weak lensing on the background number counts goes as $N_{\rm obs}/N_{\rm true} = 1/\mu^{(\beta-1)}$ where $\mu$ is the magnification due to the host galaxy and $\beta$ is the power-law slope of the faint end of the background galaxy flux distribution (see Equation 111 of Part 1 of Schneider, Kochanek and Wambsganss 2006). In the case of $\mz$, $\beta$ is fairly close to one (about 0.7) and thus we do not expect weak lensing to have a significant impact on our object detection.  

\begin{figure}[h!]
\centering
\includegraphics[scale=0.4]{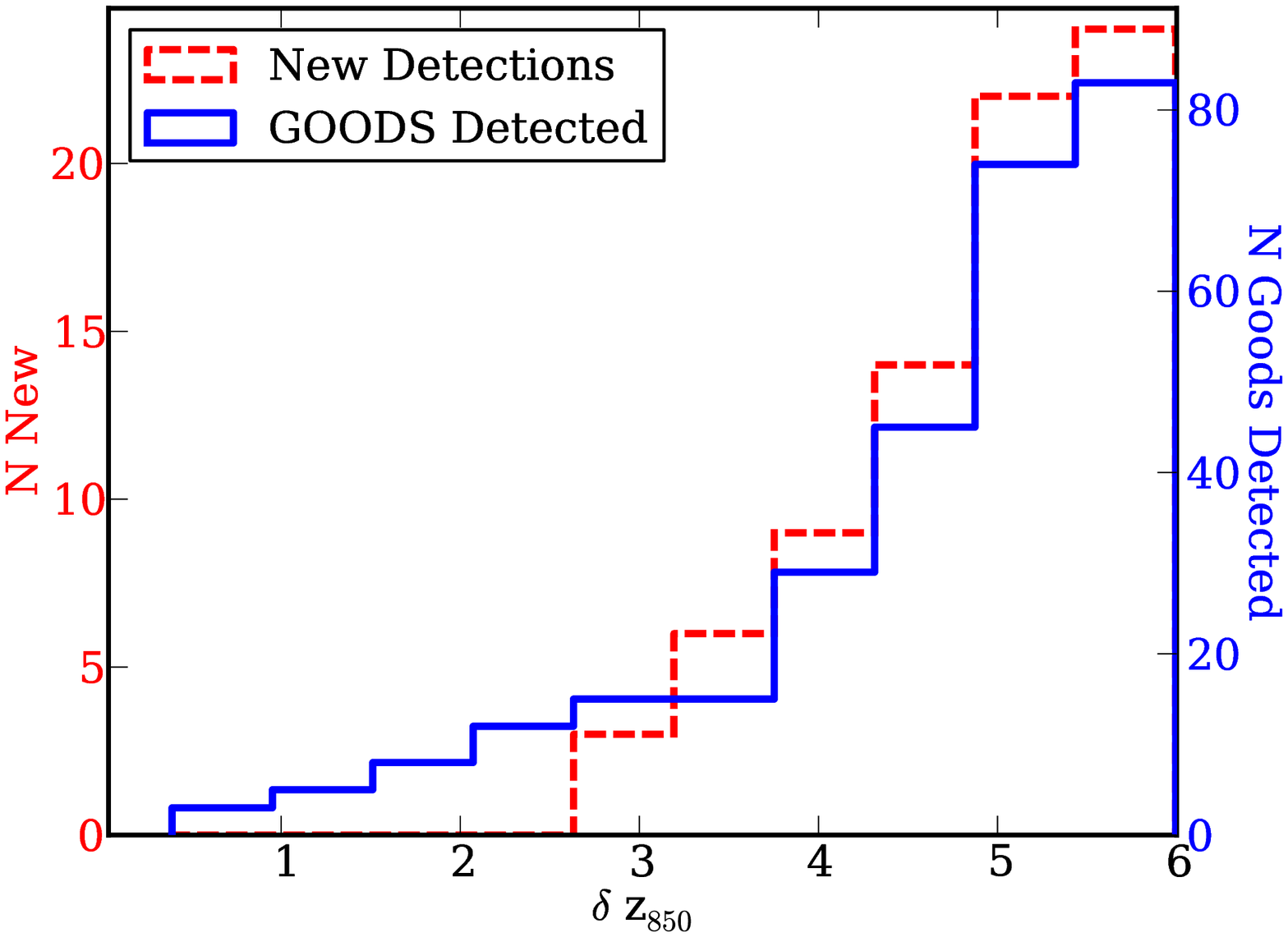}\\
\includegraphics[scale=0.4]{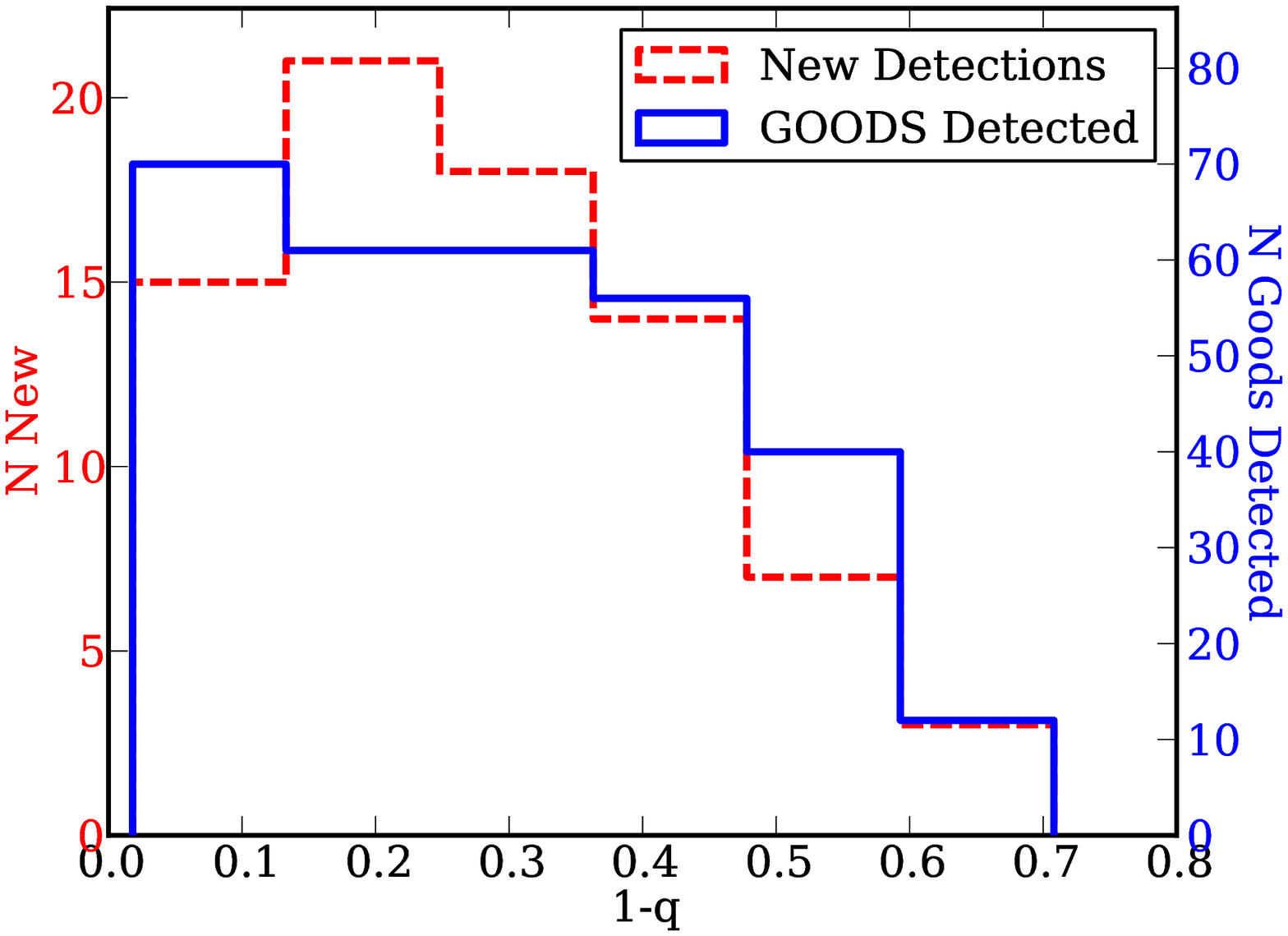}\\
 \includegraphics[scale=0.4]{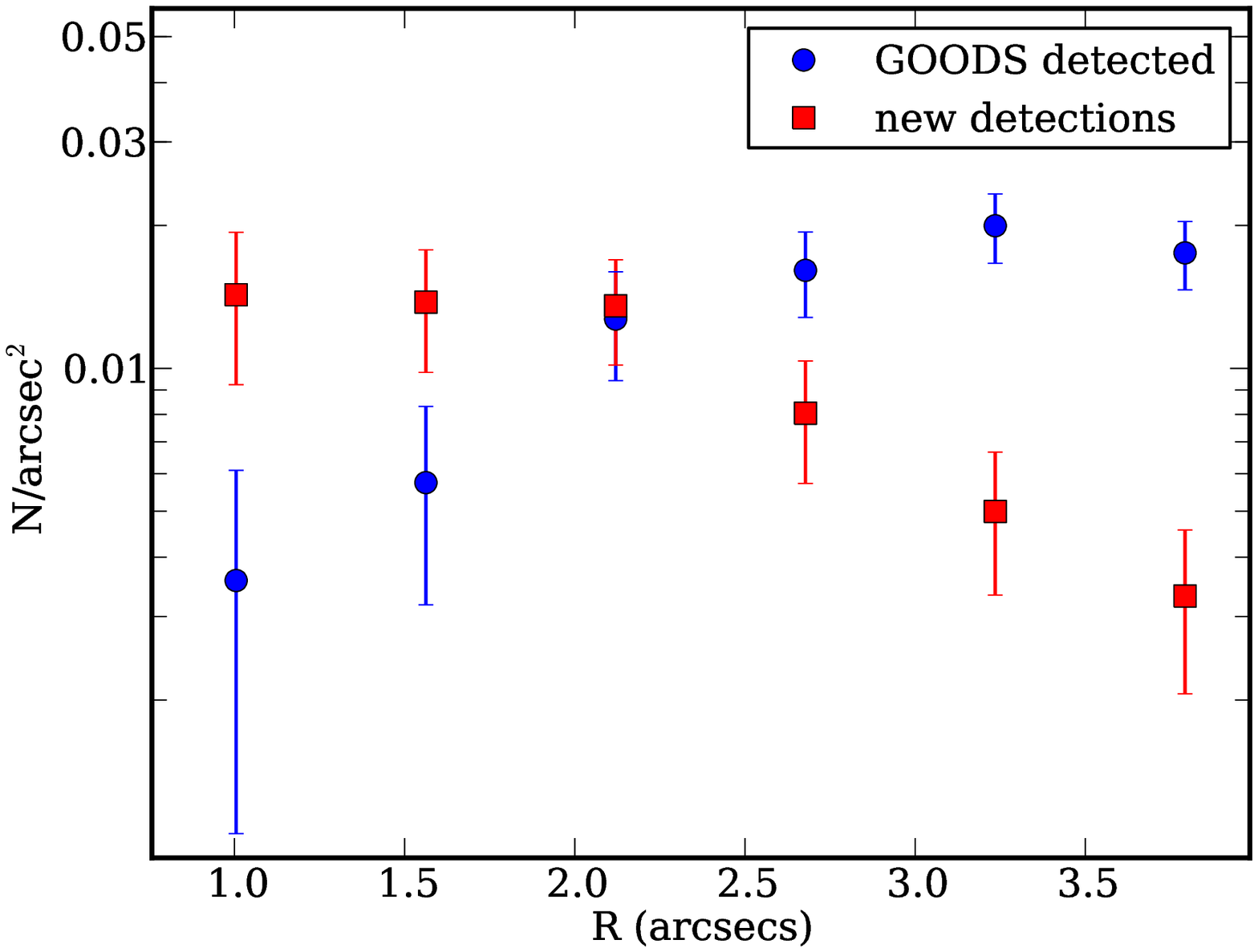}
\caption{Comparison of the properties of objects detected in the GOODS catalogs to those of newly detected objects. \emph{Top}: The distribution of object ellipticities ($1-q = 1-b/a$). \emph{Middle}: The distribution of magnitude differences
from hosts ($\delta$m$ = m - m_h$). \emph{Bottom}: Number density of objects as a function of distance from the host. Newly detected objects are closer to the host than those in the GOODS catalogs.}
\label{fig:mus}
\end{figure}

In the top panel of Figure \ref{fig:mus}, we compare the distributions
of the axis ratios of the GOODS catalog sources and the newly detected objects near
the host galaxies. The two distributions are
indistinguishable, with a Komogorov-Smirnoff (KS) probability of being
drawn from the same distribution of 0.96. The second panel of Figure \ref{fig:mus} shows the distribution of
contrast in MAG\_AUTO ($\delta$m$ = m - m_{\rm h}$) between host and
detected objects. Note the use of lower-case $\delta$, which denotes a
specific contrast from the host and is different from $\dm$ which
describes the allowed maximum contrast between host and neighboring
objects for a particular data set.  The KS value for the two
distributions being the same is $0.14$.  This value is low, but not
low enough to rule out the possibility that the distributions are the
same. Note that our improved host galaxy light subtraction procedure is important for detecting companion objects fainter than about $\delta$m$=2.5$ magnitudes (which is about 0.5 magnitudes fainter than the magnitude contrast between the Milky Way and the Large Magellanic Cloud). 

As expected, the number density radial profiles are very different for
newly detected objects and objects that had already been detected in GOODS
(bottom panel of Figure~\ref{fig:mus}), with a KS probability of $6.0
\cdot 10^{-25}$ for new and GOODS objects being drawn from the same
spatial distribution.  The number density of new detections increases
steadily towards the center of the host, while the number density of
objects already in the GOODS catalogs decreases.  Host light
subtraction triples the number density of detected objects in the
inner 3\arcsec.

\section{First Look}
\label{sec:firstlook}
The host light subtraction discussed in Section \ref{sec:subtraction}
allows us to create an enhanced catalog of objects near the host
galaxies. In this section, we show the radial and angular profiles of
objects in spatial bins in order to provide a
qualitative sense of the properties of objects near the hosts. Binning
is useful because it provides a visual representation of
data. However, it is inherently limited because it requires data
points to be averaged, thereby losing valuable information. Thus we do
not perform our analysis on the binned data but instead use this
section to justify our model choices in Section \ref{sec:model}.

\subsection{Radial Distribution}
\label{sec:BinnedRadii} 

\begin{figure}[h!]
\centering
\includegraphics[scale = 0.4]{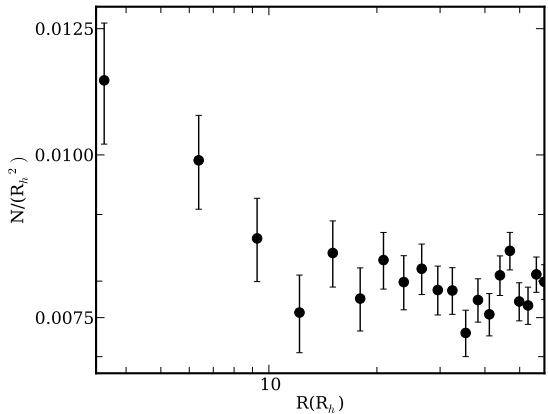}
\caption{Average number density of objects as a function of distance from the host in units of the second order moment of the host intensity profile along the major axis ($\Rhost$).}
\label{fig:prs}
\end{figure}

In Figure \ref{fig:prs} we show the average number density of objects
as a function of distance from the hosts. The number
density of sources increases dramatically near the hosts. At large radii
the number density becomes dominated by the isotropic and
homogeneous distribution of objects not associated with the hosts. In
Section \ref{sec:model} we will describe how we analyze the number density
signal by inferring the combined properties of the satellite and background/foreground 
populations.

\subsection{Angular Distribution}

\begin{figure}[h!]
\centering
\includegraphics[scale = 0.4]{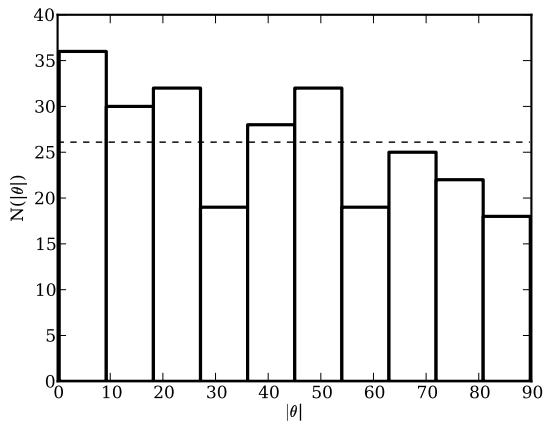} 
\caption{ Number of objects at angle $\theta$ where $\theta = 0$ is aligned with the host major axis within radius of 20 $\Rhost$ for hosts with axis ratio less than 0.6. The dashed line shows the average bin heights for a uniform distribution. } 
\label{fig:thetaDist}
\end{figure}

In Figure \ref{fig:thetaDist} we plot the angular distribution of
objects within 20 $\Rhost$ of the host galaxies, where
$\phi = 0$ is aligned with the host major axes. We show the
distribution of $|\phi|$ only for hosts with q$<0.6$ in order to
ensure all object angles are well measured (for round hosts it
becomes more difficult to measure the host position angle). The figure
shows that objects appear with more frequency towards $\phi = 0$ than
would be expected for a uniform distribution (shown by a dashed line). 
We also compare the observed angular distribution of objects to a uniform distribution by applying
a KS test which rules out the angular distribution being uniform with 95 \% 
confidence. Recall that this is \emph{without} any kind of attempt to
separate the background signal from the satellite signal. We investigate this asymmetry 
more rigorously in Section \ref{sec:model}.

\section{Joint Modeling of Satellite and Background Galaxy Populations}
\label{sec:model}

\begin{figure*}
\centering
\includegraphics[scale=0.3]{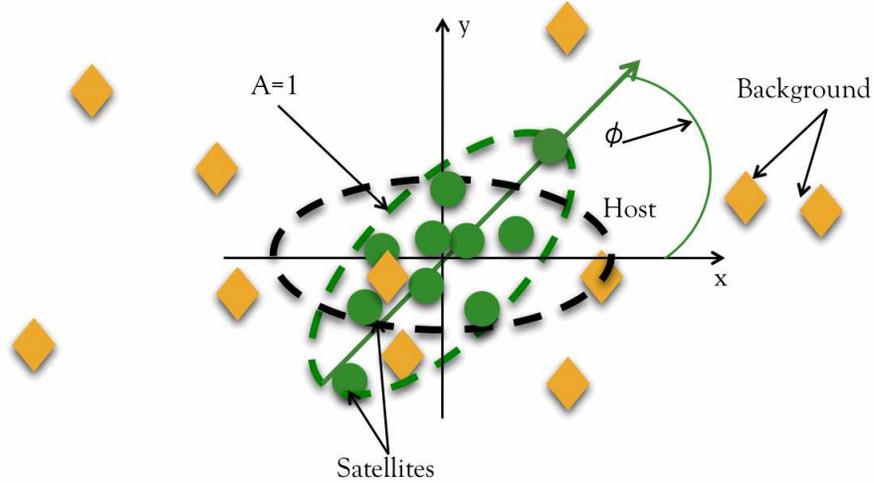} 
\caption{Schematic illustrating a possible realization of our model.}
\label{fig:schem}
\end{figure*}

We have shown that the neighboring galaxies to the GOODS host galaxies exhibit non-trivial radial and angular distributions, and we now seek to model these distributions. We start in \S~\ref{sec:satmodel} by defining the satellite model
and the parameters we aim to
constrain. In~\S~\ref{sec:backgroundmodel} we discuss the background
model and the priors we will be using. In~\S~\ref{sec:analysis} we summarize our inference
methodology, which is described in more detail in Appendix \ref{app:inference}.
A summary of all model parameters and their priors is given in Table
~\ref{table:priors}. Our implementation of the model and inference
algorithm has been extensively tested by means of simulated
data. To guide the reader, a schematic of a possible realization of
our satellite plus background model is shown in
Figure~\ref{fig:schem}.

\subsection{Satellites}
\label{sec:satmodel}
The scale-free
nature of $\Lambda$CDM results motivates us to construct our model for
the spatial distribution of the satellite number density as a function
of corresponding host parameters. For
simplicity, and owing to the relatively low signal-to-noise ratio of
our measurement, we do not allow for intrinsic scatter in these
scaling functions. Thus our inferences connecting host and satellite
properties are averages for the entire
population; individual systems will differ from the average.

We divide our discussion of our model for the satellite population
into three subsections. We describe the probability of finding a
satellite at a certain position in Subsection
\ref{subsec:spatDist}. We discuss choosing a region in which to search
for satellites in order to normalize the spatial distribution in
Subsection \ref{subsec:spatNorm}. Finally the probability of finding
a certain number of objects around each host is discussed in
Subsection \ref{subsec:satNums}.

\subsubsection{Satellite Spatial Distribution}
\label{subsec:spatDist}
$\Lambda$CDM simulations indicate that the number density of satellites 
follows the mass density profile of their host galaxy (Kravtsov 2010). In
turn, observations indicate that the three dimensional total mass
density profile of elliptical galaxies can be approximated by a power
law $\rho^{-\gamma'}$ with $\gamma'\approx2$, (with $<$10\% scatter)
\citep[e.g.][]{Koopmans++09,Aug++10}. The power law profile seems to extend 
as far as 100 $R_{\rm e}$ \citep[\eg][]{Gav++07,Lagattuta++10} which corresponds to
approximately 70 $\Rhost$. We will thus model the radial distribution of
satellites as a power law, with projected logarithmic slope
$\rpower=1-\gamma'$.


We relate the angular light profile of the hosts to the mass profile following
 the results from the Sloan Lens ACS (SLACS) by \citet[][hereafter B08]{Bol++08b}.
B08 used gravitational lensing to study the relationship between the mass and light distributions of massive elliptical galaxies.  
B08 found that the major axes of the
light profiles of the  galaxies they studied were well aligned with the total central mass profiles and that the axis ratios of the light and mass profiles were
the same within measurement errors; note that this agreement was established within the Einstein radii of the lensed galaxies, which corresponds to roughly $\sim$ 1\arcsec \citep[see also][]{Koc2002}. The majority of our
hosts are well represented by the properties of the SLACS lenses which
are massive elliptical galaxies with stellar masses in the range $\sim
10^{10.5}$ to $10^{12} M\sun$ \citep{Aug++09}\footnote{We assume that
the small number of hosts ($\sim 10$) in our sample which were less
massive do not deviate significantly from the relationship between mass and light
orientation observed in their more massive counterparts.}.  
This implies that we can expect the angular distribution of satellites
to roughly follow that of host light and therefore
that it is reasonable to construct a model of the satellite
angular profile which is related to parameters which describe the host
light angular profile.

We model the satellite distribution in
an elliptical coordinate system described by an angle $\theta$ which is the
same as the normal polar angle, and a radial coordinate $R'$ related
to Cartesian coordinates x, y by:
\be
R' = \sqrt{x^2 + y^2/q^2}
\ee

Where $q$ is the ratio between minor and major axes of the elliptical coordinate system.
The probability of finding a satellite on an elliptical contour
within the area element $R'dR'd\theta'$, goes as a power law with
power $\rpower$
\be
P_{\rm s}(R')R'dR'd\theta' \propto (x^2+y^2/q^2)^{\rpower/2}dxdy.
\ee

Changing to polar coordinates gives
\begin{equation}
\begin{array}{l}
P_{\rm s}(r,\theta|\phi,q,\rpower) \propto \\
\\
r^{\rpower}[\cos^2{(\theta - \phi)} + 1/q^2\sin^2{(\theta - \phi)}]^{\rpower/2}rdrd\theta
\label{eqn:satelliteSpatDist}
\end{array}
\end{equation}
where we introduce $\phi$ to allow for some offset between the major
axis of the satellite elliptical coordinate system and the major axis of the host galaxy; $\phi = 0^{o}$ corresponds to parallel
alignment, and $\phi = 90^{o}$ corresponds to perpendicular alignment
(see Figure \ref{fig:schem}). This is different from the coordinate
$\theta$, which describes the offset between a particular object and
the host major axis. We are only interested in the magnitude of the
offset between the host light profile and the satellite population so
we infer $|\phi|$, the absolute value of the offset of the
distribution from the host major axis.

In general the offset $\phi$ in 3 dimensions is related to the 2D
projected offset in a non-trivial way because of the random
orientation of the host-satellite system in the plane perpendicular to
the projection plane. However, for many interesting and plausible
scenarios, the alignment in 3 dimensions relates in an obvious way to
the observed, projected offset $\phi$.  Namely, if the satellites are
aligned with the host light distribution in 3D then the projection of
the two systems will also appear aligned. Similarly, if the systems
are anti-aligned, the satellites will appear perpendicular to the host
in projection. Finally, if the satellites are oriented randomly with
respect to the host in 3 dimensions, their projection will appear
isotropic. Thus the projected relationship between the satellite
spatial distribution and the host light profile contains relevant
information about the 3D distribution.

We also aim to determine the connection between the ellipticity of the
host light and that of the satellite distribution. For this purpose,
we define the ellipticity to be:
\be
\epsilon = \frac{1-q^2}{1+q^2},
\ee
We introduce a parameter $A$ to relate the ellipticity of the host
profile, $\epsh$, to that of the satellite distribution, $\epss$:
\be
\epss = \frac{\epsh A}{1 + \epsh (A-1)}
\ee
We choose this parametrization because it returns valid ellipticities
($0<\epsilon<1$) for all values of $A>0$. This is convenient when
exploring the $A$ parameter space with our MCMC code.  The parameter
$A$ can be understood as follows: when $A = 0$, the satellite
distribution is always round regardless of how flat the host
distribution is; when $A=1$ the satellite distribution has the same
flattening as the host distribution; and for values of $A>1$, the
satellite distribution is more flat than the host light distribution.
Note that formally $A$ goes from zero to infinity. As $A$ approaches
infinity, the satellite distribution approaches $q = 0$.  We are
interested in a qualitative characterization of the satellite
flattening so we simplify our inference on $A$ by restricting our
analysis to $0<A<2$, which will allow us to distinguish between
isotropic and flattened satellite distributions without having to
explore an unnecessarily large parameter space.

\subsubsection{Spatial Normalization}
\label{subsec:spatNorm}
We choose a maximum and minimum
radius in which to search for satellites in order to normalize our distribution. These radii are determined by observational constraints. The inner radius is constrained to
be $r_{\rm min} = 1.5 \Rhost$ by our
ability to accurately measure faint object magnitudes near the host (see Section \ref{subsec:objectDetection}). For the outer radius we choose $r_{\rm
max} = 45 \Rhost$ which corresponds to approximately  140 kpc. This
choice is a compromise between studying an area large enough to find
LMC/SMC equivalent objects (which are $\sim$ 60 kpc from the Milky Way), and
keeping the area small enough to limit overlap between host galaxies
and allowing us to reasonably apply a background prior determined by
the entire field.

The normalized radial probability distribution is thus:
\be
P_{\rm s}(r|\rpower) = \left(\frac{\rpower + 2}{r_{\rm max}^{\rpower +2}-r_{\rm min}^{\rpower +2}}\right)r^{\rpower +1},
\ee
The normalization of the angular distribution is given by a
generalized elliptical integral.

As discussed in Section \ref{subsec:objectDetection}, a few of our
systems are not complete to the same inner radius due to 
issues with light modeling for highly flattened
hosts. Furthermore, some of the regions far from the hosts are not
complete to 45 $\Rhost$ due to GOODS field edge effects. We discuss
how we account for this incompleteness in Appendix
\ref{app:completeness}.
 
\subsubsection{Number of satellites per host}
\label{subsec:satNums}

Multiple dark matter simulations have found that the number
of satellites with a given mass relative to the host mass is constant
\citep{Moore++1999,Krav++04,Gao++04}.
With this in mind, we model the number of satellites within a fixed \emph{magnitude} range
from the host  as being drawn from a Poisson distribution with some constant mean $N_{\rm s}$. 

It is important to keep in mind the following caveats. Due to baryonic
physics, the observable properties of galaxies are not exactly scale
invariant. In fact, the virial mass to light ratio is not a universal
function and we expect it to vary for the host galaxies and 
satellites.  This means that in a fixed magnitude range we are not
actually probing a fixed virial mass range, but only a fixed range in
luminosity ratio. However, it should be noted that our host galaxies
are typically luminous enough (several $10^{10} L_\odot$) that their
satellites are also significantly brighter than the typical Local
Group dwarf galaxies where the M/L is believed to be much higher than
in massive ellipticals. Thus our satellite mass range is better characterized by the
relative luminosity between host and satellites than it would be for the Local Group. 

Naturally, these assumptions will have no effect on our inference of
the parameters of the satellite spatial distribution, provided that it
is also independent of host galaxy luminosity within the spatial and mass ranges
considered here.


\subsection{Background/Foreground Objects}
\label{sec:backgroundmodel}
In addition to satellites, each of the host galaxies is surrounded in
projection by background/foreground objects. We isolate the satellite
signal by using the properties of the entire GOODS fields to provide
a strong constraint on the background number density. 

Galaxies tend to cluster on scales of 
order a typical galaxy virial radius, or $\sim 400$ kpc \citep[e.g.][]{Totsuji++1969,Peebles++1974,Brainerd++1995,Villumsen++1997,Zehavi++02,Morganson++09}. This clustering is believed to be due to the accretion of matter along dark matter filaments \citep[e.g.][and references therein]{Benson++2010}. Because of 
clustering, the density of objects tends to be higher in regions near bright galaxies and to have fluctuation amplitudes larger than Poissonian. \citet{Chen++06} tested a variety of methods for removing `interloper' (background/foreground) contamination from their satellite signal using a set of $\Lambda$CDM simulations. \citet{Chen++06} found that simply estimating 
the number density of background objects by studying randomly selected 
regions in their field significantly underestimated the contamination from foreground/background objects that appeared near their hosts in projection.
They found the most reliable estimate of the background was obtained by measuring the background in annuli centered on their hosts and just outside of the area in which they studied the satellite population. 
We adopt this method to build a prior on the background near the hosts. We study the background in annuli between 45 and 60 $\Rhost$ (typically 140-180 kpc). We choose this distance range to ensure that we are not including satellites in our estimate, while still accurately characterizing the clustering of galaxies near the hosts. We find that the density of objects near the hosts with magnitudes within our detection range ($21<\mz<26.5$) is $\Sigma_{b,o} = 125\pm 2$ arcmin$^{-2}$.  As expected, this is higher than the average density of objects in the GOODS fields ($117 \pm 0.6$) and has larger fluctuations than one would predict from Poisson noise. 

Recall that we are studying objects brighter than a fixed magnitude
contrast from the host magnitudes. This means that the
number of objects we study around a given host is a fraction of the density of 
objects brighter than $\mz = 26.5$. We correct for this around
each of the hosts by representing  the cumulative distribution function (CDF) of the background number counts by a power-law \citep[e.g.][]{Benitez++03}
\be
N_b(<m_{\rm max}) \propto 10^{\alpha_{\rm b} m_{\rm max}}
\ee
where $\alpha_{\rm b}$ is the slope of the background number counts.  
The maximum magnitude $m_{\rm max}$ for a particular host system is
\be
m_{\rm max} = m_{\rm h}+\dm.
\ee
Thus for a given choice of $\dm$, the expected number density of
background/foreground objects around the $j^{\rm th}$ host is
\be
\Sigma_{\rm b, j} = \Sigma_{\rm b,o}
10^{\alpha_{\rm b}(m_{\rm{h}, j} + \dm - m_{\rm lim})}.
\ee
We measure the background slope to be $\alpha_{\rm b}= 0.28 \pm 0.01$ in the same annuli near the hosts in which we estimate the number density of the background.

\subsection{Analysis}
\label{sec:analysis}
In the previous two subsections we constructed a model which
describes the probability of a satellite or foreground/background
object appearing a given distance from the host dependent on a choice
of parameters. The parameters and their priors are listed
in~Table~\ref{table:priors}. Of key importance in our work is that our results are unbinned and
analyzed in a fully Bayesian fashion. This means that for each parameter
our result is a posterior probability distribution function (PDF) which describes
the probability of a value of a parameter being true given our data
(the likelihood) and our prior knowledge of the parameter (the prior).

In Appendix \ref{app:inference} we discuss the details of constructing
the posterior probability function (Equation \ref{eqn:fullLikelihood})
which we use to study our parameter posterior PDFs. We compute the posterior PDFs using a Markov Chain Monte Carlo (MCMC) method. At
least $10^6$ iterations per chain are performed in order to ensure
convergence.

\begin{deluxetable*}{clc}[h!]
\tabletypesize{\small}
\tablecaption{\label{table:priors}
Summary of model parameters}
\startdata
\hline
Parameter        & Description & Prior \\
\hline
Satellite Model  &             & \\
Ns                    &   Number of satellites per host & U(0,20) \tablenotemark{a}\\
$\rpower$        & Logarithmic slope of the satellite radial distribution & U(-5,0)\\
$A$       & Flattening of the satellite number density distribution relative to host light flattening & U(0,2) \\
$|\phi|$  & Offset between the major axis of satellite spatial distribution and the major axis of the host light profile. & U(0,$\pi/2$)\\
\hline
Background Model & & \\
$\Sigma_{\rm b,o}$       & Number density of background objects with magnitudes between $21 < \mz<26.5$& N(125,2) \tablenotemark{b} \\
$\alpha_b$        & Logarithmic slope of the $\mz$ background number counts & N(0.28,0.01)\\
\enddata
\tablenotetext{a}{U(a,b) denotes a uniform distribution between a and b.}
\tablenotetext{b}{N($\mu$,$\sigma$) denotes a normal distribution with mean $\mu$ and
standard deviation $\sigma$.} 

\end{deluxetable*}


\section{Results}
\label{sec:results}
\begin{figure*}[h!]
\centering
\includegraphics[scale=.8]{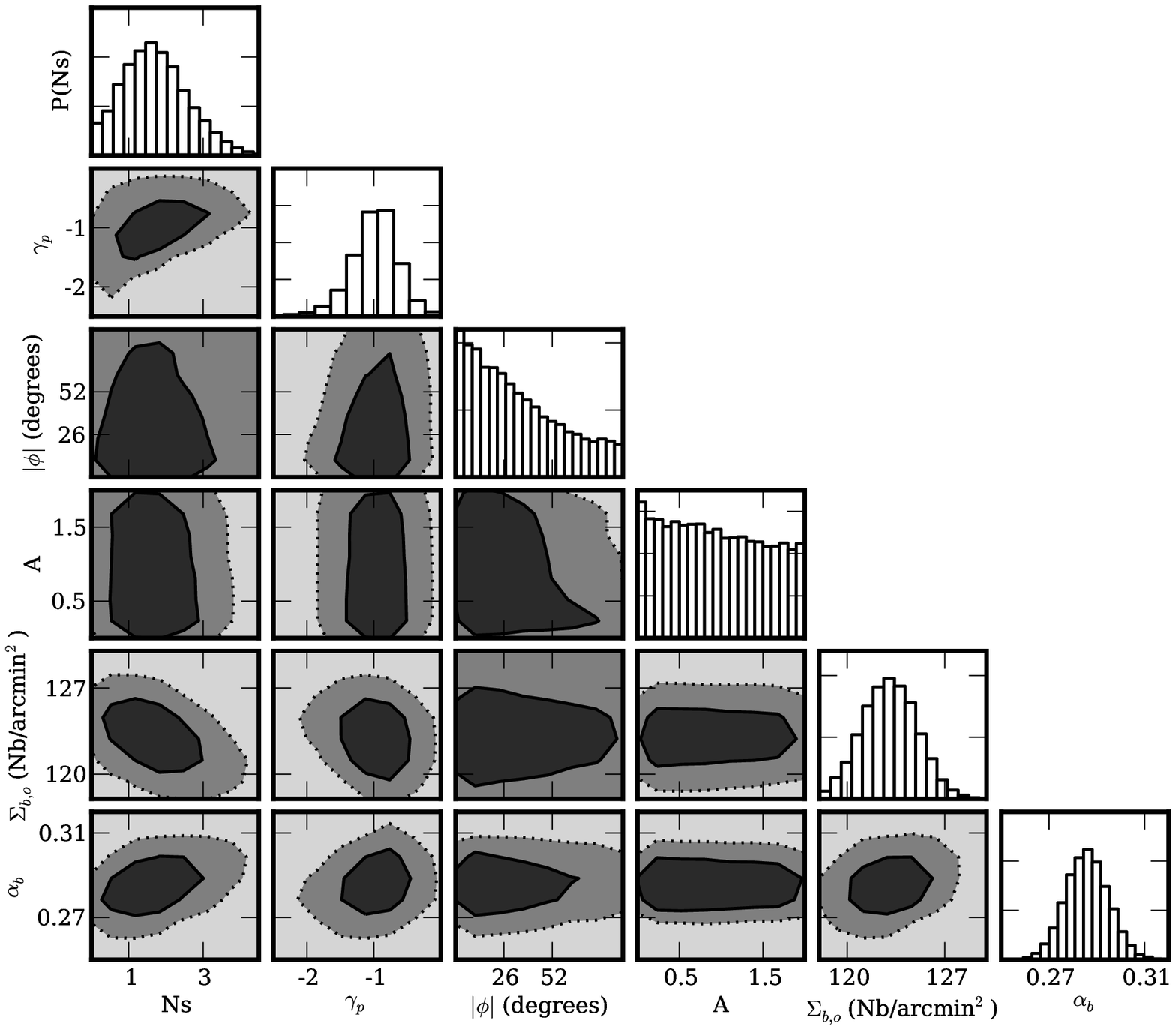} 
\caption{Bivariate posterior PDFs for all model parameters for the $\dm = 5.5$ data set. The dark and light
contours contain regions of 68 and 95\% of the probability
respectively. The diagonal shows the marginalized PDF for each
parameter.}
\label{fig:mcdm55}
\end{figure*}

We first discuss the
results for the $\dm = 5.5$ magnitude bin, which has the strongest
signal.  Table~\ref{table:results} contains a summary of the results
of the inference. The posterior PDF for each variable is shown in
Figure~\ref{fig:mcdm55}.

The first main result is the clear detection of a peak in the posterior PDF for
$\Numsat$ which is well removed from zero, indicating the detection of a
population of satellites. Secondly, we find that the radial density
profile of the satellite spatial distribution is consistent with isothermal
($\rpower=-1$). Thirdly, the posterior PDF of $\phi$ is
peaked at zero indicating that the satellite spatial distribution is
preferentially aligned with that of the host. A KS test shows that a
uniform distribution of angles is ruled out at more than a 99.99\% CL.

\begin{figure*}[h!]
\centering
\includegraphics[scale=.8]{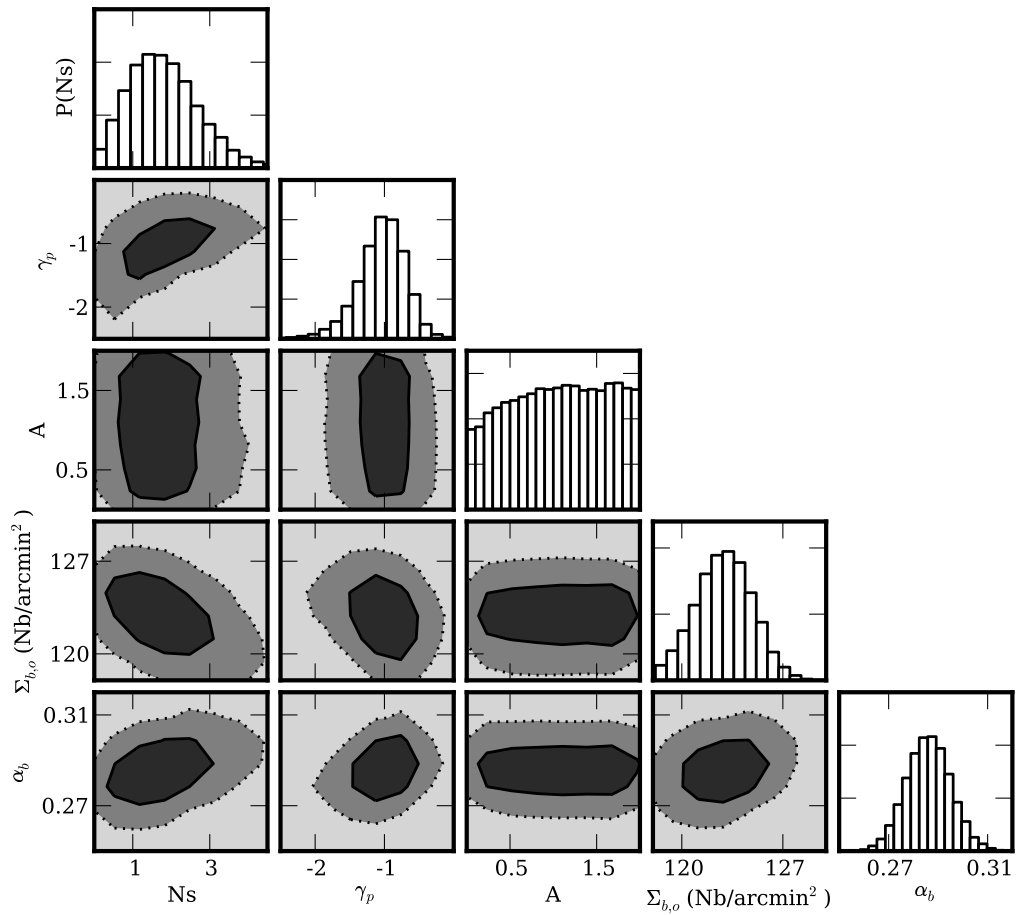} 
\caption{Same as Figure~\ref{fig:mcdm55}, but with $\phi$ fixed at 0.}
\label{fig:fixedPhi}
\end{figure*}

Our results are inconclusive for the relationship between ellipticity
of the satellite distribution and that of the host, described by the
parameter $A$ (see Figure \ref{fig:mcdm55}).  This is not surprising
as it takes a stronger signal to measure the ellipticity of a
distribution than to measure its alignment.  Our inability to infer
$A$ is in part due to a degeneracy between $\phi$ and $A$. If $A$
is zero, all values of $\phi$ are equally probable. This can be seen in the
contour plot shown in Figure \ref{fig:mcdm55}.  We show the effects of
removing this degeneracy by performing a separate analysis on the
$\dm=5.5$ data set, keeping $\phi$ fixed at zero. The results for this
are shown in Figure \ref{fig:fixedPhi}. When $\phi$ is fixed at zero,
the inference disfavors small values of A, which implies a disfavoring
of satellite distributions that are rounder that that of the stars.

We also verified that our results are robust to small changes in
limiting magnitude by repeating the analysis for $\dm = 6.0$ and 5.0
magnitude bins.  The $\dm = 6.0$ bin clearly shows the presence of a
satellite population, with the same isothermal slope observed for the
$\dm = 5.5$ range and the same angular alignment with the
host major axis. The inferred satellite number is also consistent with
the $\dm = 5.5$ measurement, although there is a longer tail towards
higher satellite numbers which might indicate a slightly higher
number, as expected given the fainter limit. As discussed in Section~\ref{sec:sample}, the
number of hosts that are complete almost halves from $\dm = 5.5$ to
$6.0$ so the errors are larger for the $\dm = 6.0$ inference
overall. Interestingly, the analysis for $\dm = 5.0$ did not
conclusively detect a satellite population; the number of satellites is
marginally more than 1-$\sigma$ greater than zero, and the uncertainties on the parameters
describing the spatial distribution are similarly larger. This
shows the crucial need for deep data in performing this kind of
measurement. In all cases we recover our priors on the background
parameters $\alpha_b$ and $\Sigma_{\rm b,o}$. We are not able to
constrain these numbers further because in our model they are both
degenerate with each other and with the number of satellites.

\begin{figure*}[h!]
\centering
\includegraphics[scale=.8]{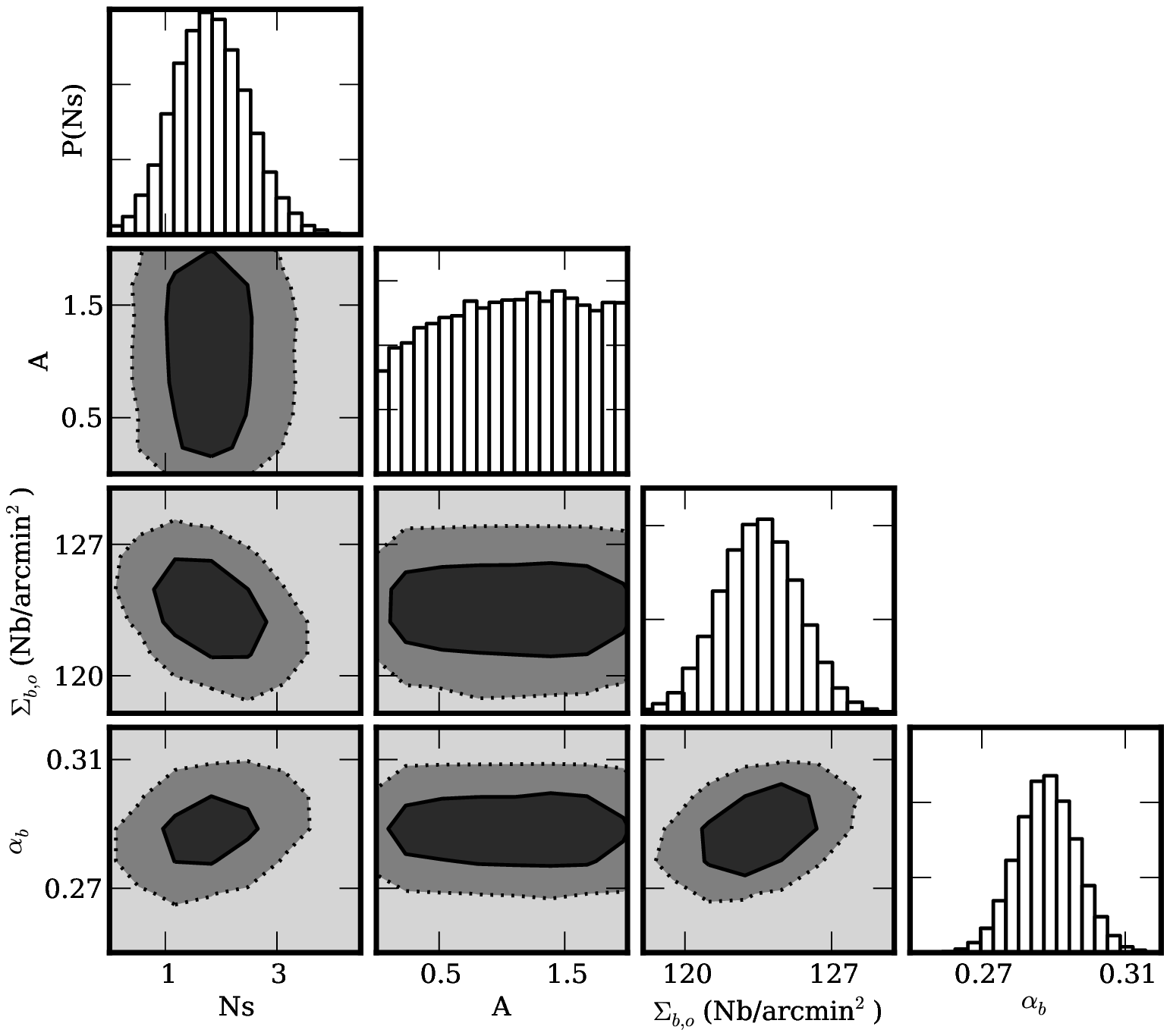} 
\caption{Same as Figure \ref{fig:mcdm55}, but with fixed $\phi = 0$ and $\rpower=-1$.}
\label{fig:mcdm55_fixedPhiG}
\end{figure*}

An important degeneracy is that between $\numsat$ and $\rpower$. The
inferred number of satellites is larger for a shallow radial profile and
vice-versa. From theoretical and observational arguments (see the
Introduction) we expect the radial profile to be close to isothermal and
we expect alignment between host and satellites. It is thus instructive to
repeat the analysis by fixing these parameters to their expected
values $\rpower=-1$ and $\phi=0$, thereby eliminating many of the
degeneracies. The resulting posterior PDFs for $\dm=5.5$ are shown in
Figure~\ref{fig:mcdm55_fixedPhiG}, while results for all magnitude ranges
are summarized in Table~\ref{table:fixedPhi}.

As expected, fixing $\rpower$ and $\phi$ lowers the uncertainty of
our inference and the detection of satellites becomes more
significant. It is also easier to compare the results across magnitude
bins, and we see how the number of satellites indeed increases with
$\dm$ and is consistent with a cumulative luminosity function going as
$L^{-1}$ (i.e. similar to the cumulative mass function predicted from
theory assuming $m \propto L^{-2}$), albeit with large
errors. Furthermore, the posterior PDF of $A$ begins to disfavor a
satellite population that is more isotropic than the stars in the
galaxy. This is also broadly in line with expectations, as we expect
that the distribution of stars has been made rounder than the host halo by
dissipational processes.

\begin{deluxetable*}{clllllll}[h!]
\tabletypesize{\small}
\tablecaption{\label{table:results}
Posterior Medians/Confidence Intervals}
\startdata
\hline
$\Delta m$ & $\rpower$& Ns & $|\phi|$&A& $\Sigma_{\rm b,o}$(N$_{\rm b}/$arcmin$^{2}$) &$\alpha_b$  \\
 &  &                                       & (68\% confidence) &    & \\
\hline \\
6.0 & $-1.1^{+0.5}_{-0.6} $&$3^{+1}_{-1}$& $ <44$& -\tablenotemark{a}& $123^{+2}_{-2}$&$ 0.283^{+0.009}_{-0.009}$ \\
5.5 & $-1.0^{+0.3}_{-0.4} $&$1.7^{+0.9}_{-0.8}$& $ <42$&- & $123^{+2}_{-2}$&$0.286^{+0.009}_{-0.009}$ \\
5.0 & $-0.5^{+0.4}_{-0.8} $& $0.5^{+0.8}_{-0.4}$&$<56$& -& $124^{+2}_{-2}$&$0.282^{+0.009}_{-0.008}$ \\
\enddata
\tablenotetext{a}{No inference on the parameter because the posterior distribution is approximately uniform.}
\end{deluxetable*}

\begin{deluxetable*}{clllllllc}[h!]
\tabletypesize{\small}
\tablecaption{\label{table:fixedPhi}
Posterior Medians/Confidence Intervals with fixed parameters}
\startdata
\hline
Fixed parameter&$\Delta m$ & $\rpower$& Ns &A& $\Sigma_{\rm b,o}$(N$_{\rm b}/$arcmin$^{2}$) &$\alpha_b$  \\ 
& &                      &      &(68\% confidence) & & \\
\hline\\
$\phi = 0$& 5.5 & $-1.0^{+0.3}_{-0.4} $&$1.7^{+1.0}_{-0.8}$& $>0.73$& 
$123^{+2}_{-2}$&$0.286^{+.009}_{-0.009}$ \\
\hline \\
$\phi = 0$,  $\rpower = -1$& 6.0 & -&$2^{+1}_{-1}$& $>0.83$& 
$124^{+2}_{-2}$&$0.285^{+0.009}_{-0.009}$ \\
 $\phi = 0$,  $\rpower = -1$& 5.5 & -&$1.8^{+0.7}_{-0.6}$& $>0.72$& 
$124^{+2}_{-2}$&$0.288^{+0.008}_{-0.008}$ \\
$\phi = 0$,  $\rpower = -1$& 5.0 & -&$0.4^{+0.4}_{-0.3}$& $>0.73$ & 
$124^{+2}_{-2}$&$0.281^{+0.008}_{-0.008}$ \\

\enddata
\end{deluxetable*}

                      
\section{Discussion}
\label{sec:discussion}

This is the first measurement of the numbers and spatial distribution
of faint satellites of early-type galaxies at intermediate redshift so
it is useful to provide some context for our result.  Our measurement
of the radial slope of the satellite spatial distribution is
consistent with an isothermal distribution, i.e. $\gamma_{\rm p} \approx
-1$. This result is similar to that found in W10, and the radial
distribution of satellites appears to be consistent with that of the
total mass distribution measured in lensing studies
\citep[e.g.,][]{Koopmans++09, Aug++10}. However, given the uncertainty on the
inferred slope, the spatial distribution of satellites is also
consistent with the NFW profile inferred by C08. In the radial range
covered by our work ($\sim$5--140 kpc), an NFW profile is
characterized by a radially averaged projected logarithmic slope of
approximately $\gamma_{\rm p} \sim -1$, with large variations
depending on host scale radius. More accurate measurements of the
radial density profile of satellites are needed, in addition to a
comparison of the radial profile around different host masses, in
order to determine whether the distribution of satellites follows
that of the total mass or that of the dark matter component.

The distribution of satellites is anisotropic and is
preferentially aligned along the major axis of the host light
profile. This host-satellite alignment is consistent with observations
of host-satellite systems in
SDSS~\citep[e.g.][]{Brainerd++05}. Furthermore, alignment is predicted
by $\Lambda$CDM simulations which show satellites accreting
anisotropically along filaments and appearing aligned with the major
axis of the host mass profile \citep[e.g.,][]{Aubert++04}.
Previous observations have established that the host light profile aligns
with the
host mass profile (B08), and our observation of the satellite-host light
alignment therefore implies alignment between the satellite distribution
and the host mass, in agreement with simulations. This alignment between host
mass and satellite spatial distribution has important implications for
the frequency of flux ratio anomalies, as we discuss below.

\begin{figure}[h!]
\centering
\includegraphics[scale=.5]{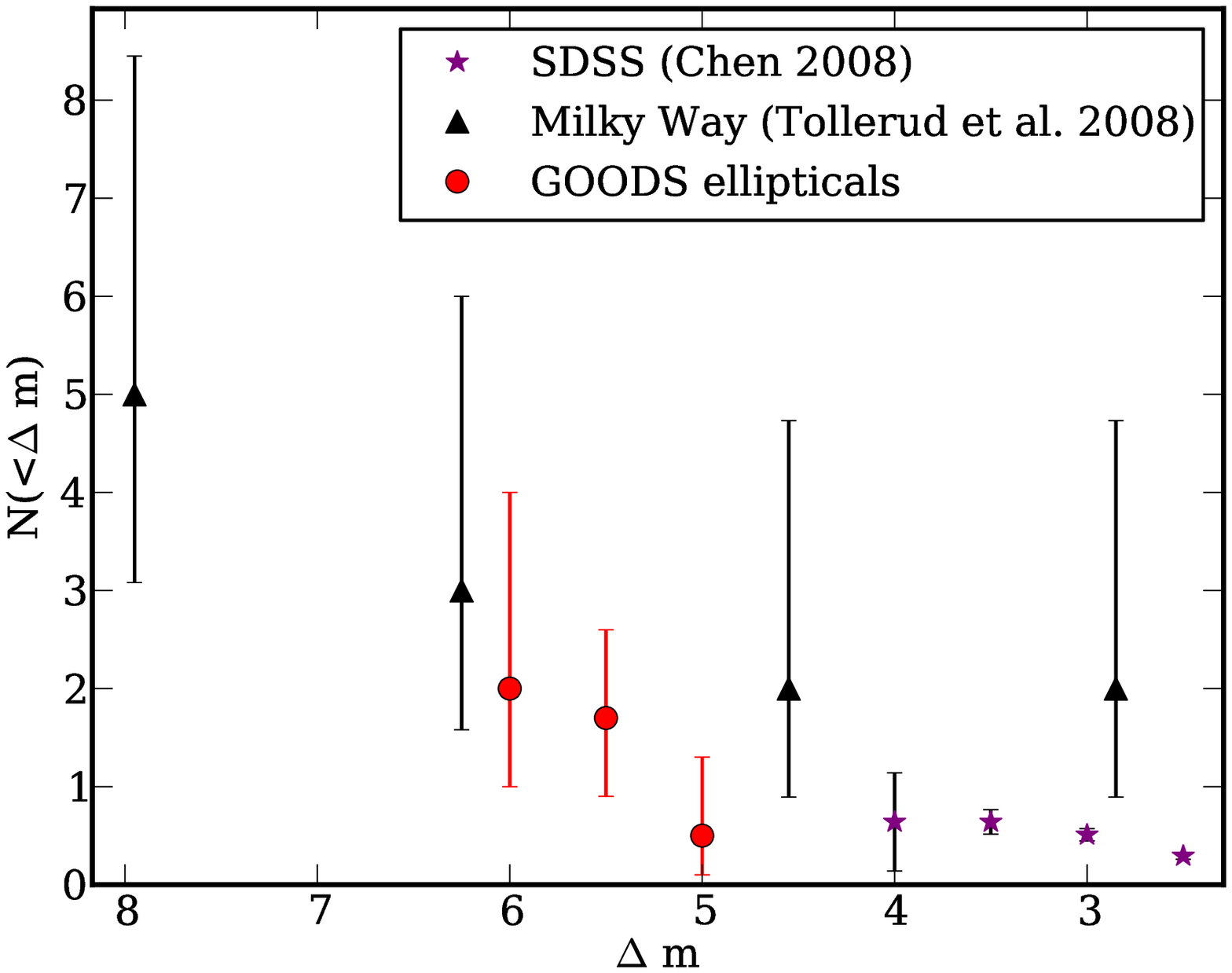} 
\caption{Comparison of the measured number of GOODS satellites to those found for the Milky Way and SDSS galaxies.}
\label{fig:satLF}
\end{figure}

Although we do not expect the number of satellites of intermediate
redshift elliptical galaxies to be exactly equal to that of galaxies
at lower redshift due to baryonic processes and evolution, it
is instructive to compare the GOODS satellites to other satellite
populations.  Figure~\ref{fig:satLF} compares the cumulative
luminosity function (CLF) of our satellites to the CLF of the
Milky Way and Andromeda satellites~\citep[adopted from][]{Toll++08}, with one
sigma uncertainties calculated from
\citet{Gehrels++1986}, and to the CLF of SDSS satellites of hosts
with varying morphologies adopted from C08. Ideally, we would have
also liked to compare our study to J10 which is one of the few studies
of satellites of high redshift objects. However, the maximum value of
$\dm$ in that survey was about 2.5 magnitudes which is too bright to
be compared to our measurement in a meaningful way.

C08 measured the number of satellites and their radial profile between
approximately 20 and 350 kpc.  In order to compare our results
appropriately, we use the C08 `interloper subtracted' fit to a radial
power law given in their Table 2 in order to extrapolate their
 measurement inward toward the host. We assume that the power law is constant in the
inner regions and use a fiducial value of $\Rhost \sim 3$ kpc to
estimate the number of satellites C08 would have seen in the region
that we studied (i.e., as close as 1.5~$\Rhost$). The Milky Way CLF is complete in the same equivalent
region as ours so we did not have to make any adjustments in order to
compare our numbers.  All three studies measured magnitudes in
different filters; C08 used $r$ band magnitudes at an effective redshift of $z = 0.1$, we use observed-frame $\mz$ magnitudes, and
Milky Way measurements are in rest-frame $V$. However, the corrections required to convert from $r$ at redshift $\sim 0.1$ and observed $\mz$ at redshifts
of $\sim 0.5$ into rest-frame V are negligible compared to the sizes of
the bins of $\dm$ that we study, and we therefore do not include explicit $k$-corrections in our analysis. The three satellite CLFs are roughly consistent where the measurements overlap, given the relatively large error bars (Figure 11). It is worth noting that our GOODS satellite measurements have significantly smaller error bars than those for the Local Group by virtue of the large sample of hosts we study which allows us to reduce the sampling error. Furthermore, we are able
to observe much fainter satellites than the SDSS study.

It is also instructive to compare our inferred satellite number to the
number of minor mergers our hosts are expected to undergo in the time
span we study. As many of our satellites are very near their host
galaxies, we expect that some of them will be close to merging and
that our estimate of the satellite number should account for the
predicted minor merger rate in the mass range we study.  The merger
rate depends on the ratio between the host and satellite virial masses
\citep[e.g.,][]{Fakhouri++10}. Our typical host has a stellar mass of M*
$\sim 10^{10.5}$ M$\sun$ (see Figure \ref{fig:host_redshifts}). From
abundance matching techniques \citep[e.g.,][]{Beh++10}, this
corresponds to a halo mass of about $10^{12}~{\rm M}\sun$. In the luminosity
range of our satellites the stellar mass to
light ratio of the satellites should be approximately in the range
30-100\% of that of the host galaxy \citep[e.g.,][]{Kauff++03}. Using this, the stellar mass of the satellites we study is in the range of $0.6-2\cdot10^{8}$
M$_{\odot}$. This corresponds to virial masses of $\sim
2-8\cdot10^{10}$ M$_{\odot}$, according to the best fit to Equation 21
in~\citet{Beh++10}. Thus the we can detect satellites with virial
masses of order a percent of that of their hosts. This estimate is
subject to large uncertainties, including those arising from tidal
forces that tend to remove dark matter from satellites as they move
closer to their host galaxy, thus reducing their virial mass.  Using
the best fit to Equation 1 in~\citet{Fakhouri++10}, we estimate that the
majority of our hosts (60\%) will undergo about 0.4-0.5 mergers
(depending on the satellite stellar mass to luminosity fraction)
between redshifts 0.1 and 0.8. It is reassuring that this number can
be easily accounted for by our inferred satellite number. Of course, a
quantitative comparison requires a detailed evaluation of the merging and
visibility timescales, which is beyond the scope of this
paper. However, we will re-visit this point in a future paper, as 
minor mergers have been suggested to play an important role in
the evolution of early-type galaxies
\citep[e.g.][]{Bundy++07,BoylanKolchin++08,
Naab++09,Nipoti++09,Kaviraj++09,Hopkins++10,Kaviraj++11}.

Finally we discuss the effects our detected satellite
population would have on flux ratio anomalies in quadruply-imaged
strong gravitational lens systems. We limit our discussion here to order of magnitude
estimates and general trends, leaving a rigorous quantitative analysis
to a future paper.

\citet[][hereafter D02]{Dalal++02} used the magnifications of images in five quad lenses
to infer a mass profile for the lens galaxies. In order to reproduce the observed
magnifications, D02 inferred the 
presence of a non-smooth component to the mass distribution with a mass fraction between 0.6 and 7 \% near the lensed images.
\citet{Mao++04} and more recently \citet[][hereafter X09]{Xu++09} used $\Lambda$CDM simulations to test whether dark substructure embedded within smooth halos could produce the flux ratio anomalies observed by D02. These studies estimated that the mass fraction of CDM substructure from their simulations would only be about 0.1 \% near image positions, which is insufficient to produce the flux ratios studied by D02. The inclusion of globular cluster populations and baryonic processes does not change this result \citep{Xu++10} and underscores the importance of constraining the satellite population through direct observations. 

From our inference, a few percent of our hosts have a satellite within
the annulus that would include the typical lens image positions studied by X09 (the Einstein radius is approximately 1\arcsec).
We therefore find that the average mass fraction in this annulus
is of order a few percent. Using the mass function ($dn/dm
\propto m^{-1.9}$) described by X09, we estimate that the average
satellite mass fraction they would have observed in our mass range is
of order a percent, which is consistent with our result. It is worth
considering the possibility that the presence of a few relatively
massive satellites may increase the lensing cross-section
significantly so that any attempt to measure the satellite mass
function using gravitational lenses would be automatically biased
towards very high mass fractions near the lens Einstein radius; we will explore this possibility in more detail in paper III.

It is interesting to note that, although our typical hosts have similar
halo masses to those studied in X09, the most massive subhalo X09
observed in any of their simulations has a virial mass of $\sim 10^{10}~{\rm M}~\sun$, which
is on the very low mass end of the subhalos studied in this work. Our
simple extension of the virial to stellar mass relationship led us to
estimate that our hosts will have on average more than one halo \emph{at
least} this massive even assuming that the mass to light ratio of our
satellites is only a few percent of that of the host halo.  This
discrepancy may be due to the fact that our mass estimate is very
approximate and, as already discussed, may not be appropriate for
systems undergoing strong interactions such as tidal stripping.  On
the other hand, the difference in estimated satellite masses may
indicate the importance of baryonic condensation in preserving the
mass in subhalos during accretion \citep[see e.g.,][]{Weinberg++08,
Romano++09}.

Our observation that satellites appear preferentially along the host
major axis may offer an additional route to reduce or eliminate the
apparent discrepancy in the satellite mass fraction as inferred by
flux ratio anomalies.  As pointed out by \citet{Zentner++06}, an
anisotropic satellite distribution can increase the effective
projected mass of the host `felt' by a lensed image by as much as a
factor of six.  Thus, the mass associated with our observed luminous
satellites could in principle be sufficient to explain the anomalies
detected by D02.

Our satellites are much fainter than a typical lensed quasar image and
thus would be easily obscured in any of these lensing studies. Our
result highlights the importance of studying the satellite population
in non-lens systems in conjunction with the analysis of perturbations
in lens systems in order to attain an understanding of the mass and
luminosity functions of satellites.

To conclude, it is important to point out that the largest
discrepancies between the luminosity and mass function are seen for even
fainter satellites than the ones probed by this study
\citep{Krav++10}. It is thus desirable to push the analysis of both the
luminosity and the mass function of satellites to even smaller
fractions of the host galaxy properties.

                      
\section{Summary}
\label{sec:summary}

We use an advanced host light subtraction method to study
the spatial distribution of faint galaxies around 127 early-type galaxies in
the GOODS fields between redshifts 0.1 and 0.8. We employ a
self-consistent model in the framework of Bayesian inference to
disentangle the satellite population from background/foreground
galaxies. Exploiting the depth and resolution of the HST images, we
detect satellites up to 5.5 magnitudes fainter than the host 
galaxy and as close as 0\farcs5/2.5~kpc to the host. Our main
results can be summarized as follows:

\begin{enumerate}
  \item Intermediate redshift, massive, early-type galaxies have on
  average $1.7^{+0.9}_{-0.8}$ satellites within our luminosity
  ratio limit. This is consistent with the number of satellites
  observed in the Milky Way.  

\item The number density of satellites follows an
  approximately isothermal radial power law profile $P(r)\propto
  r^{\rpower}$ with $\rpower = -1.0^{+0.3}_{-0.4}$. 

\item The satellites are preferentially aligned along the
  major axis of the host light profile with $|\phi|<42^ {o}$ at the
  68\% confidence level.

\item When the offset $\phi$ between the satellite population and host light profile is fixed at 0 degrees, the satellite distribution
is inferred to be preferentially more elongated than the distribution
of host galaxy light.

\item When the satellite number density profile is assumed to be isothermal, the average number of satellites is inferred to be 1.8$^{+0.7}_{-0.6}$.

\end{enumerate}


\acknowledgments

We thank N.~Jackson, L.V.E.~ Koopmans, R.~Wechsler, M.~Busha, P.~Schneider, D.D.~Xu,  for many insightful comments and stimulating conversations. We thank M. Giavalisco and the rest of the GOODS team for their work on the GOODS ACS images and catalogs. We thank J.~Chen for providing data for Figure \ref{fig:satLF}. AMN and TT acknowledge support
by the NSF through CAREER award NSF-0642621, and by the Packard
Foundation through a Packard Fellowship.  PJM was given support by the
Kavli Foundations and the Royal Society in the form of research fellowships.


\bibliographystyle{apj}
\bibliography{references}

                      
\appendix
             
\section{Host Galaxy Light Subtraction}
\label{app:subtraction}

As we showed in Section~\ref{sec:subtraction}, the detectability of satellite
galaxies is quite sensitive to the presence of light from the much brighter
host galaxy. In this appendix we give more details on how the light from the
host galaxy was subtracted in order to allow the faint satellites to be
detected.
Host galaxy surface brightness modeling is a two step process. In the
first step, we identify and mask all objects other than the host in the region
in which we want to model host galaxy light. This ensures that our host model
does not attempt to fit the light of nearby objects. 
Once the objects are identified and the mask is created, in the second step 
we model the host
galaxy surface brightness in 2D 
using the unmasked parts of the image, interpolate over
the masked parts, and subtract this model from our original image.

\subsection{Masking}

The masking is done automatically using the segmentation map produced by an
initial \sextractor run, with parameters tuned
to optimize the detection of small faint objects in the presence of the host
light.
This optimization was performed by
simulating faint companion objects at varying positions around
our hosts.  We found that using the GOODS \sextractor parameters at this stage
gave unsatisfactory results, in that objects we simulated near
the hosts were not being masked. 
We found that, in order to
properly account for both faint point sources near our hosts and diffuse
sources further from our hosts, we needed to 
use a superposition of masks created with
two different sets of \sextractor parameters. These parameters are 
listed in Table~\ref{table:SePars}. 
With this masking routine, we were able to 
accurately recover magnitude 27 point
sources as close as 1.5~$\Rhost$ (typically $\sim$ 0\farcs9) from 
our host galaxy centroids.

\begin{deluxetable}{cccc}[h!]
\tabletypesize{\small}
\tablecaption{\label{table:SePars}
\sextractor Parameters}
\renewcommand{\arraystretch}{1.5}
\startdata
\hline\hline   
Parameter        & \multicolumn{2}{c}{Value}                   \\
                 & Large Object Mask & Point Object Mask& Final Photometry     \\
\hline
DETECT\_MINAREA  & 10                & 5        &      --      \\
DEBLEND\_NTHRESH & 64                & --      &      --       \\
DEBLEND\_MINCONT & 0.001             & --     &     --         \\
DETECT\_THRESH   & --                & 2.5         &     --    \\
ANALYSIS\_THRESH & --                & 2.5         &     --    \\
FILTER\_NAME     & --                & gauss\_2.0\_3x3.conv & --\\
BACK\_TYPE       & MANUAL            & MANUAL    & MANUAL          \\
BACK\_VALUE      & 0.0               & 0.0                &  0.0
\enddata
\tablecomments{Parameters not listed, or marked as ``--'', are
those used by the GOODS team and can be found at.
\texttt{http://archive.stsci.edu/pub/hlsp/goods/catalog\_r2}}
\end{deluxetable}

\subsection{B-spline model subtraction}

In Figure~\ref{fig:modExamples} we show some representative examples
of our host galaxy light subtraction process.  Some of our hosts are
very elongated or show disk-like features. The surface brightness distribution of
these objects is more difficult to model and tends to leave
residuals at larger radii than the rounder hosts. Residuals are easily
identifiable because of the symmetric pattern and distinct elongated
shape that they appear in (see, e.g., the residuals of the upper right
host in Figure~\ref{fig:modExamples}).  We deal with this small
number ($\sim 20$) of more disky/extended sources in two ways. The
first way is to increase the amount of flexibility in the B-spline
fit, increasing the number of multipoles fitted in each of our spline
rings around the ellipse. The second thing we do is to exclude a
larger inner region from our analysis to ensure we are not
identifying any residuals as objects. 

\section{Source Extractor Parameters and Photometry Comparison}
\label{app:SE}

After subtracting the main galaxy light out of the image we run
\sextractor on the residual image to identify the remaining
objects. In this step we modify our
\sextractor parameters by increasing the deblending threshold in order to match the
parameters used when making the GOODS catalogs.  A full list of our \sextractor
parameters is listed in Table \ref{table:SePars}.

Even after matching the deblending parameters, there are two minor
differences between our final \sextractor parameters and GOODS
\sextractor parameters. The first is that GOODS estimates the
background by measuring the noise in an annulus with 100 pixel (3\arcsec) width around each object.  This method of background
subtraction is not optimal for our measurement because we are focusing
on a relatively small area (typical image size was $\sim200$x$200$
pixels), within which we expect to have a relatively high object
density. Instead, we estimate the background by calculating the 3
$\sigma$ clipped mean within our image.
Furthermore, the GOODS team made modifications to \sextractor that
changed the deblending process slightly.

\begin{figure}[h!]
\centering
\includegraphics[scale=0.4]{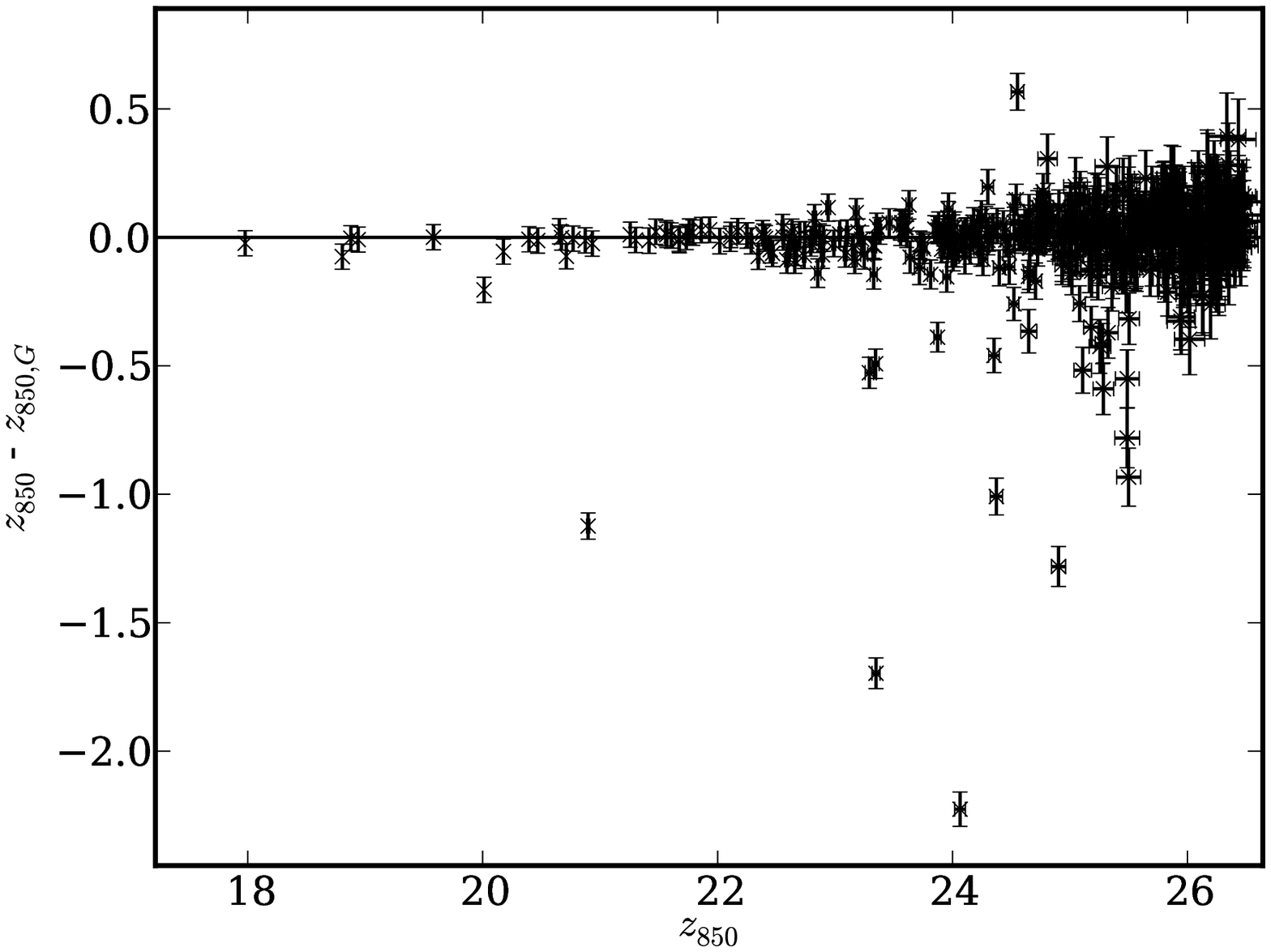} 
\caption{Comparison of our photometry with GOODS in a 4 arcmin$^2$ field cutout with no host galaxy subtraction.}
\label{fig:fieldphot_magauto}
\end{figure}

We compare the MAG\_AUTO output from \sextractor for the two methods in
a 4 arcmin$^2$ cutout from the GOODS field in order to study the
effects of the different background subtraction and deblending
(see Figure \ref{fig:fieldphot_magauto}). Using our parameters, we identify 484
objects with MAG\_AUTO$<$26.5.  Of these, 22 objects do not have a center
within 0\farcs3 of an object in the GOODS catalog. The major
outliers in the MAG\_AUTO comparison are all in areas of high object
density and thus most likely due to differences in deblending. The
mean difference (not including major outliers) in the MAG\_AUTO estimate is
$(-1.9 \pm 6.0)\cdot 10^{-3} $.

\section{Completeness}
\label{app:completeness}
In Section \ref{sec:model} we discuss our model for the spatial distribution of satellites which we can use to calculate the probability that there are a certain number of objects $\hat{\numobjs}$ within $1.5<r/R_h<45$, with positions $\pos$. Before comparing a particular model prediction to the data it is necessary to account for observational limitations such as field edge effects. 
In this section we explain how we account for these observational limitations when inferring our model parameters.

Following \citet{Kelly++07}, we define a completeness function, $P(I| \pos)$,
which describes the probability of observing an object at position $\pos$, where $I = 1$ indicates an object is detected and $I = 0$ indicates a non-detection. Note that because we have chosen to study only magnitude ranges in which we are complete, our completeness function will only depend on position in a given field. As discussed in Appendix \ref{app:subtraction}, we measure our completeness near the host galaxy by placing simulated objects around the host and calculating our recovery rate and the accuracy of our photometry as a function of position. Far from the host, we estimate our completeness in annuli to account for field edge effects. To simplify this analysis, we ignore regions of partial completeness so that a region in the field is either 100\% complete or 0\% complete.

Taking completeness into account, the likelihood of detecting
$\numobs$ total objects (satellites and background),with positions
$\{\pos\}$, given model parameters $\modpars$, becomes
\be
\prob(\nobs,\{\pos\}|\modpars) = \prob(I|\{\pos\})\prob(\hat{\nobjs},\{\pos\}|\modpars)
\ee

Our fields have varying sizes but the model parameter $\hat{\numobjs}$
gives the number of objects we expect to find in an ideal field where
we have 100\% completeness between $1.5$ and $45$ $\Rhost$. As discussed above, some of our
host systems have restricted areas, between $r_{mn}$ and $r_{mx}$,
where the total area is smaller than in the ideal case. In these cases we must scale the model prediction
by the probability of finding the object in the reduced area compared
to the probability of finding the object in the ideal area, and we therefore define an updated prediction for the number of objects $\hat{\nobjs}'$ given by
\be
\hat{\nobjs}' = \hat{\nobjs}\frac{\int_{r_{mn}}^{r_{mx}}\prob(\pos'|\modpars)d\pos'}{\int_{1.5}^{45}\prob(\pos'|\modpars)d\pos'}
\ee
Note that $\hat{\numobjs}$ is a smooth model prediction of the number
of true objects in a field. In order to properly account for the fact
that the true number of objects in a field is discrete, we must
introduce a Poisson probability, which relates the ``true'' discrete
number of objects to the model number of objects.
The likelihood that we observe $\numobs$ objects at positions $\pos$  given our model parameters, is a product over the likelihood of each of those positions being true given our model parameters 
\be
\prob(\{\pos\}|\modpars) = \prod_i^{\numobs}\prob(\pos_i|\modpars)
\ee
Thus the probability that we observe $\nobs$ given a set of model parameters, taking into account varying total 
areas of completeness, is given by:
\be
\prob(\nobs,\{\pos\}|\modpars)= \prob(\numobs|\hat{\nobjs}')
\prod_i^{\numobs}\prob(\pos_i|\modpars).
\ee
Properly, we should also include a term which accounts for the probability of \emph{not observing} $\numobjs - \numobs$ objects. However, in our case, the region of parameter space we exclude due to observational limitations
is always much smaller than the region we do study, and thus the number of objects our model predicts that we expect to observe is much larger than the number that we cannot observe. Therefore, for simplicity, we do not include this term in our analysis.
\section{Inference methodology}
\label{app:inference}
In this section we discuss in detail how we characterize the posterior PDFs for
our satellite model parameters $\modpars_{\rm s} = \{N_s,\rpower, A,\phi\}$, and our background model parameters $\modpars_{\rm b} = \{\Sigma_{\rm b,o}, \alpha_b\}$. We present a top-down description of our model in which we start with the posterior PDF we would like to obtain and break it apart to show
where we have inserted different pieces of information. 

Our final data set, $\allDat$, is composed of a set of measurements for all of our host-field systems, $\allDat = \{\allDat_1,\allDat_2,..., \allDat_{N_h}\}$
Where $\allDat_{j} = \{\hdat_j, \data_j\}$, and $\hdat_j$ is the set of 
measurements of the magnitude, axis ratio and RMS of the $j^{th}$ host light profile, 
and $\data_j = \left[\numobs_j,\{(\pos_1),(\pos_2),...(\pos_{\numobs_j})\}\right]$ is the number of objects observed
around the $j^{th}$ host and their positions.
Using Bayes' theorem we can express the probability of a set of model parameters being true given the data by:
\be
\prob(\modpars|\bs{D}) \propto \prob(\allDat|\modpars)\prob(\modpars)
\label{eq:basic_fn}
\ee
In the above equation the term $\prob(\allDat|\modpars)$ is the
likelihood of the data being true given a set of model parameters and
$\prob(\modpars)$ is the prior information we have about the model
parameters. Our model parameters and their priors are listed in Table \ref{table:priors}. 

The first term on the right of Equation \ref{eq:basic_fn} is a product of likelihoods for individual host systems:
\be
\prob(\allDat|\modpars) = \prod_j^{\numhosts}\prob(\data_j|\modpars,\hdat_j)
\label{eq:moreComplicated_fn}
\ee

Notice that in our model, the measurements of the host light profile are treated as model parameters which 
we assume are known exactly. This is because we have constructed our model such that
its parameters do not 
predict the properties of the host light profile.

Equation \ref{eq:moreComplicated_fn} is itself composed of a product of the 
likelihood of measuring $\numobs_j$ objects around each host times the probability for measuring each object position,
\be
\prob(\data_j|\modpars,\hdat_j) = \prob(\numobs_j|\modpars)\prod_i\prob(\pos_i|\modpars,\hdat_j)
\ee
Recall that in Section \ref{sec:model} we built a model that was
composed of separate contributions from a satellite and background
population. This means that the probability of finding any object at a given location is the sum of the probability of finding a background object at that location with the probability of a finding a satellite at that location:
\be
\prob(\data_j|\modpars,\hdat_j) = \prob(\numobs|\modpars)\prod_i \prob(\pos_i|\modpars,\hdat_j,S)\prob(S|\modpars,\hdat_j)+\prob(\pos_i|\modpars,\hdat_j,B)\prob(B|\modpars,\hdat_j)
\label{eqn:fullLikelihood}
\ee
The term $\prob(\pos_i|\modpars,\hdat_j,S)$
is the probability of finding a satellite at a certain position (given by Equation \ref{eqn:satelliteSpatDist}) and $\prob(\pos_i|\modpars,\hdat_j,B)$ is the probability of finding a background/foreground object at a certain position (which is uniform). The term $\prob(S|\modpars,\hdat_j)$ is the probability of an object being a satellite given only the model parameters,
\be
\prob(S|\modpars,\hdat_j)= \frac{N_{s}}{N_s+N_b}
\ee
and $\prob(B|\modpars,\hdat_j)$ is defined in an analogous way for the line-of-sight interlopers. The functions that determine $N_s$ and $N_b$ in terms of model
parameters are discussed in Section \ref{sec:model}.

\end{document}